\documentclass[aps, prd, floatfix, nofootinbib, superscriptaddress, twocolumn]{revtex4-1}

\usepackage{latexsym}
\usepackage{amsmath}
\usepackage{amssymb}
\usepackage{amsfonts}
\usepackage{textcomp}
\usepackage{color}
\usepackage{multirow}

\usepackage[mathscr,scaled=1.15]{urwchancal}
\DeclareFontFamily{OT1}{pzc}{}
\DeclareFontShape{OT1}{pzc}{m}{it}%
{<-> s * [1.15] pzcmi7t}{}
\DeclareMathAlphabet{\mathpzc}{OT1}{pzc}{m}{it}

\usepackage{color}
\usepackage{supertabular}
\usepackage{placeins}
\usepackage{epsfig}
\usepackage{graphicx}

\definecolor{purple}{rgb}{0.5,0,0.5}
\definecolor{blue}{rgb}{0.0,0,0.9}
\definecolor{prdblue}{rgb}{0.133,0.118,0.498}
\usepackage[colorlinks=true, pdfstartview=FitV, linkcolor=prdblue, citecolor= prdblue, urlcolor=prdblue]{hyperref}

\usepackage{xspace}
\newcommand{\jpsi}{\ensuremath{J/\psi}\xspace}
\newcommand{\mevnospace}{\ensuremath{{\mathrm{\,Me\kern -0.1em V}}}}
\newcommand{\gevnospace}{\ensuremath{{\mathrm{\,Ge\kern -0.1em V}}}}
\newcommand{\tevnospace}{\ensuremath{{\mathrm{\,Te\kern -0.1em V}}}}
\newcommand{\mev}{\mevnospace\xspace}
\newcommand{\gev}{\gevnospace\xspace}

\renewcommand{\Re}{\operatorname{Re}}

\newcommand{\ie}{\emph{i.e.}\xspace}

\let\Im\relax
\DeclareMathOperator{\Im}{Im}
\let\Re\relax
\DeclareMathOperator{\Re}{Re}

\makeatletter

\setbox0\hbox{$\xdef\scriptratio{\strip@pt\dimexpr
    \numexpr(\sf@size*65536)/\f@size sp}$}

\newcommand{\scriptveryshortarrow}[1][3pt]{{%
    \hbox{\rule[\scriptratio\dimexpr\fontdimen22\textfont2-.2pt\relax]
               {\scriptratio\dimexpr#1\relax}{\scriptratio\dimexpr.4pt\relax}}%
   \mkern-4mu\hbox{\let\f@size\sf@size\usefont{U}{lasy}{m}{n}\symbol{41}}}}

\makeatother

\begin{document}

\title{Study for a model-independent pole determination of overlapping resonances1}

\date{\today}

\author{Daniele Binosi}
\email{binosi@ectstar.eu}
\affiliation{European Centre for Theoretical Studies in Nuclear Physics
and Related Areas; Villa Tambosi, Strada delle Tabarelle 286, I-38123 Villazzano (TN), Italy}
\author{Alessandro Pilloni}
\email{alessandro.pilloni@unime.it}
\affiliation{Dipartimento di Scienze Matematiche e Informatiche, Scienze Fisiche e Scienze della Terra,
Universit\`a degli Studi di Messina, Viale Ferdinando Stagno d'Alcontres 31, I-98166 Messina, Italy}
\affiliation{INFN Sezione di Catania, Via Santa Sofia 64, I-95123 Catania, Italy}
\author{Ralf-Arno Tripolt}
\email{Ralf-Arno.Tripolt@theo.physik.uni-giessen.de}
\affiliation{Institut f\"ur Theoretische Physik, Justus-Liebig-Universit\"at Giessen, Heinrich-Buff-Ring 16, D-35392 Giessen, Germany}
\affiliation{Helmholtz Research Academy Hesse for FAIR (HFHF), Campus Giessen, D-35392 Giessen, Germany}

\begin{abstract}
We apply a model-independent reconstruction method to experimental data in order to identify complex poles of overlapping resonances. The algorithm is based on the Schlessinger Point Method where data points are interpolated using a continued-fraction expression. Statistical uncertainties of the experimental data are propagated with resampling. In order to demonstrate the feasibility of this method, we apply it to the $S$-wave $J/\psi \to \gamma \pi^0\pi^0$ decay. We benchmark the method on known analytic models, which allows us to estimate the deviation from the true value. We then perform the pole extraction from BESIII data, and identify the $f_0(1500)$, $f_0(1710)$, and $f_0(2020)$ scalar states. Our results are in reasonable agreement with recent results, which suggests the proposed method as a promising model-independent alternative for the determination of resonance poles that is solely based on available experimental data.
\end{abstract}

\maketitle


\section{Introduction}

The discoveries of unexpected exotic resonances in the last two decades have brought new life to Hadron Spectroscopy. Multiquark, gluonic and multihadron states have drawn attention from a larger community, and hopefully a comprehensive understanding of their nature will shed light on the inner mechanisms of QCD~\cite{pdg,Esposito:2016noz,Olsen:2017bmm,Guo:2017jvc,Lebed:2016hpi,Karliner:2017qhf,Guo:2019twa,ali2019multiquark,Brambilla:2019esw}. Present experiments are providing datasets with unprecedented statistics, and we can expect further dramatic improvements in both statistics and resolution in the near future and, especially, with forthcoming facilities. It is thus necessary to develop theoretical analysis tools that are at least as refined. 

The rigorous determination of resonance parameters and properties requires to extract the poles' position and residue from a reaction amplitude. More specifically, one has to perform the analytic continuation to the complex plane of a function of which we know the modulus on the real axis only, notoriously a difficult problem. The $S$-matrix principles put strong constraints on the properties of scattering amplitudes, in particular in the elastic region; and dispersive techniques have allowed to establish the lightest scalar multiplet with remarkable precision~\cite{Caprini:2005zr,DescotesGenon:2006uk,GarciaMartin:2011nna,Hoferichter:2011wk,Moussallam:2011zg,Ditsche:2012fv,Pelaez:2020uiw,Pelaez:2020gnd}. Above the inelastic threshold, however, the strict application of the $S$-matrix constraints to coupled channels becomes quickly intractable. 

The best approach so far has remained to build amplitude models, that satisfy, at least partially, fundamental principles, and are in general motivated by some intuition of the underlying microscopic dynamics. These models depend on a finite number of parameters that can be fitted to the available data; and, if the models are analytic, one can directly continue them to the complex plane and extract the resonant poles. Some level of model dependence is thus unavoidable, and usually addressed by practitioners through exploring a large number of different parametrizations. Additionally, model artifacts can mimic actual resonant poles, and elaborate studies are needed to reasonably disentangle the two~\cite{JPAC:2021rxu}. An alternative approach is to resort to Pad\'e approximants~\cite{Masjuan:2013jha,Masjuan:2014psa,Caprini:2016uxy,Pelaez:2016klv} which are based on rational functions, whose parameters are commonly determined from derivatives of the fitted amplitude in a single point. By studying the stability of the pole position on ({\it i}) the order of the rational function, and ({\it ii}) the point at which the amplitude is calculated, one can assess a systematic error, hoping to absorb the corresponding model dependence into the amplitude parametrization.

It would be clearly advantageous to develop methods that rely on direct pole search on the available experimental data, without relying on any  assumption of the underlying physics. As their results would be inherently model independent, they would constitute a base-line against which model dependent approaches (as the ones illustrated above) can be benchmarked. One example of this type is the Laurent-Pietarinen expansion~\cite{Svarc:2013laa,Svarc:2014sqa}, which describes the reaction amplitude through a simple pole plus a background polynomial of conformal maps with its parameters fitted to data. 

\begin{figure*}[!t]
	\includegraphics[width=2.1\columnwidth]{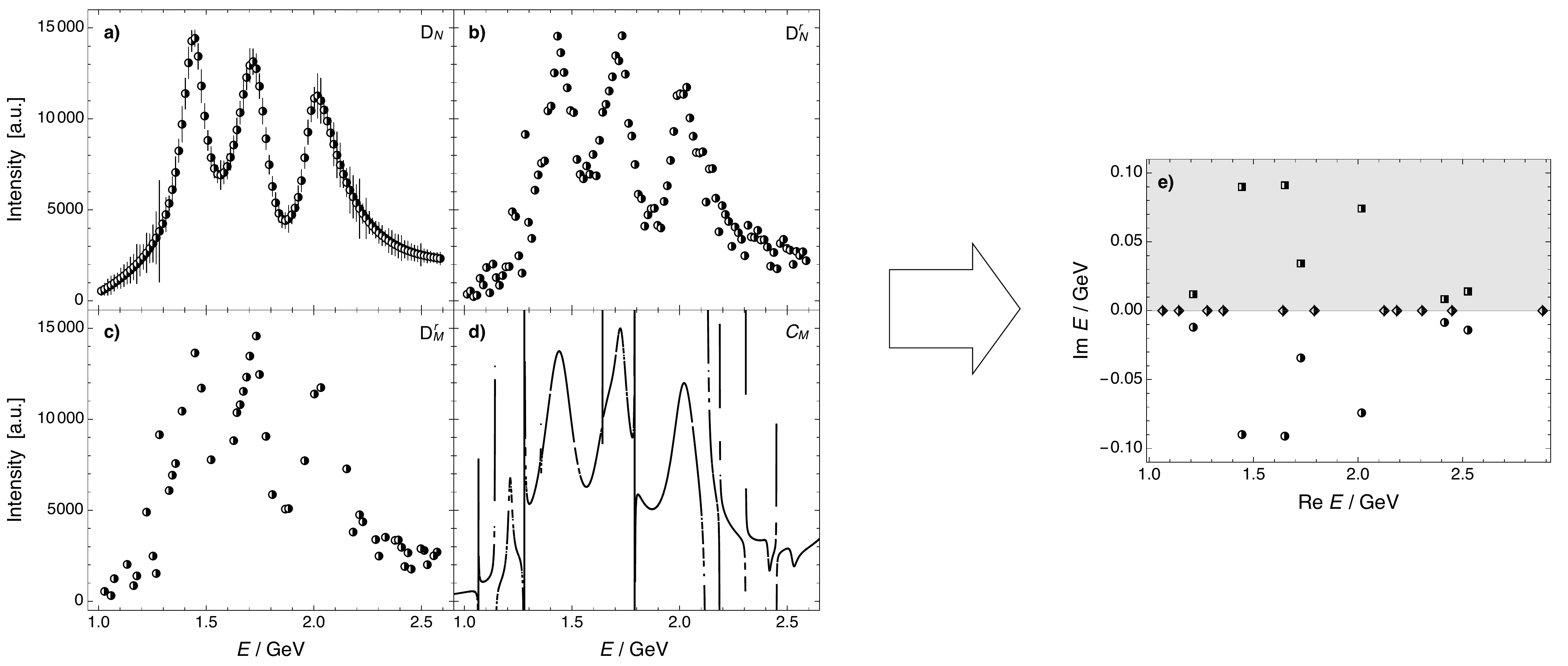}
	\caption{\label{fig:proc}{\bf SPM plus bootstrap procedure}. {\bf a)} The intensity dataset $\mathsf{D}_N$ obtained from model~A (see the description in Sect.~\ref{sec:validation} and the parameter list reported in the Appendix). {\bf b)} A typical bootstrap generated replica $\mathsf{D}^r_N$ when considering full experimental errors. {\bf c)} A randomly chosen subset $\mathsf{D}^r_M\subseteq\mathsf{D}^r_N$ ($M=50$ in the case shown); and {\bf d)} its corresponding SPM interpolator $C_M$. {\bf e)} Of $C_M$'s 25 poles, 11 are reals (diamonds) and 14 form 7 complex conjugate pairs (circles and squares), of which only the ones located in the lower part of the complex energy plane ($\Im E<0$) are considered (circles).}
\end{figure*}

In this paper we focus on the Schlessinger Point Method (SPM)~\cite{PhysRev.167.1411, Schlessinger:1966zz}, described in Sect.~\ref{sec:SPM}. Contrary to the other methods discussed, the SPM does not aim at fitting data with a given functional form, but rather {\em interpolates} the data points with a continued fraction. Having been already applied in a variety of different contexts, obtaining always remarkably accurate and precise results~\cite{Tripolt:2016cya,Tripolt:2018xeo,Chen:2018nsg,Binosi:2018rht, Binosi:2019ecz,Eichmann:2019dts,Yao:2020vef, Yao:2021pyf,Cui:2021vgm,Cui:2021aee,Cui:2021skn}, it is indeed natural to investigate its applicability to reactions of interest for hadron spectroscopy. In fact, as the SPM does not rely on fitting, it is highly affected by statistical noise, so that its performance in a realistic spectroscopy case needs to be addressed. 

To this end, the $J/\psi \to \gamma \pi^0\pi^0$ decay measured by BESIII~\cite{BESIII:2015rug} represents a good benchmark. The scalar wave above $1\gev$ features three peaks, likely associated with the $f_0(1500)$, $f_0(1710)$, and $f_0(2020)$. Together with the $f_0(1370)$, they are the main candidates for the lightest glueball, see for example \cite{Ochs:2013gi,Huber:2020ngt}. The $S$- and $D$-wave intensities and relative phase of $J/\psi \to \gamma \pi^0 \pi^0$ and $\to \gamma K_S^0 K_S^0$ were recently analyzed by JPAC with a large number of analytic parametrizations~\cite{Rodas:2021tyb}, and for each of them the pole positions have been calculated. Since all these parametrizations reproduce well the data, we can use them to produce realistic pseudo datasets, and check whether the SPM finds the right poles as a function of the statistics. This is done in Sect.~\ref{sec:validation}; and since as will be shown there, the SPM passes this validation procedure, it will be used on the real dataset; and the resonance so extracted compared with the available model results (Sect.~\ref{sec:results}).

\section{\label{sec:SPM}Statistical Schlessinger Point Method}

\begin{figure*}[!t]
	\includegraphics[width=2.08\columnwidth]{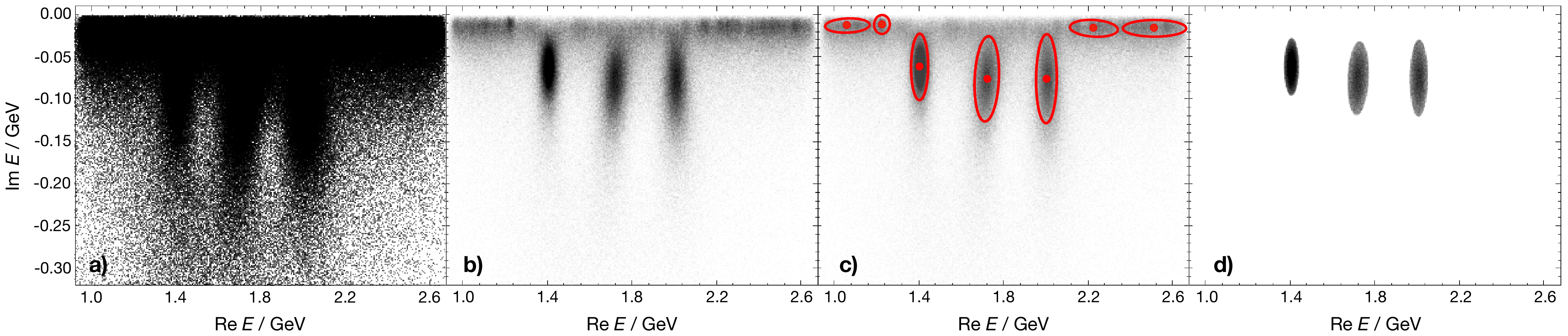}
	\caption{\label{fig:diffimg}{\bf Isolating the SPM signal from noise.} {\bf a)} Distribution of the 629,774 SPM poles obtained for model A, $M=50$,  $n_r=10^5$ and the full experimental error. {\bf b)} The use of a high transparency level for each point representing a pole present in a) isolates regions of accumulation of SPM reconstructed poles. {\bf c)} Morphological pattern recognition to extract image components representing the boundaries of identified regions: their centroid, length of the semi-axes of the best fitting ellipse, and orientation (defined as the angle between the largest axis and the horizontal axis). Noise poles clusters (if present at all) can be easily distinguished from signal ones by their small orientation angle and/or size.  {\bf d)} The extracted final clusters of  signal poles containing respectively (from left to right) 60,738, 51,666 and 46,487 poles. On average, the number of signal poles  amount to $\sim$20\% of the original poles shown in panel a).  
	}
\end{figure*}

We consider the nominal solution for the intensity of the $0^{++}\,E1$ multipole of the $J/\psi \to \gamma \pi^0\pi^0$ decay extracted from the mass-independent fit by BESIII~\cite{BESIII:2015rug}, hereafter $S$-wave. The intensity is related to the amplitude via
\begin{equation}
    I(E) = \rho(E) \left| f(E)\right|^2,
\end{equation}
where $\rho(E) = \sqrt{1-4m_\pi^2 /E^2}$ is the two-body phase space. The amplitude $f(E)$ features a right-hand cut starting at threshold, and other branch points corresponding to the opening of other channels. By analytic continuation through this cut one can access the unphysical Riemann sheet where resonant poles are found. If one is sufficiently far from threshold (as is the case in this study), one can approximate $f(E)$ with a meromorphic function, having single poles in the lower half-plane,
\begin{equation}
    f(E) \simeq \prod_P \frac{b_P E + c_P}{E - E_P}.
\end{equation}
The intensity can thus be approximated with a meromorphic function as well; and such function is real for real values of $E$, and can be written as a ratio of polynomials with real coefficients:
\begin{equation}
    I(E) \simeq \prod_P \frac{|b_P|^2 E^2 + 2E \,\Re b_p c_P^* + |c_P|^2}{E^2 - 2E\,\Re E_P+|E_P|^2}.\label{eq:intensity}
\end{equation}
Notice that in this approximation the phase space is a smooth function that can be reabsorbed in the coefficients of the two polynomials.

In what follows, we aim to study if the SPM can be used to reconstruct the analytic structure of the intensity $I(E)$ and in particular the position of (possibly all) its complex poles. 

To this end, let us denote with $\mathsf{D}_N$ the set of all available $N$ experimental pairs of the intensity $I_i$ associated to a given energy bin $E_i$:
\begin{align}
	\mathsf{D}_N = \{(E_i , I_i = I(E_i)),\ i=1,\dots, N\}.
\end{align} 
We can then select a subset $\mathsf{D}_M\subseteq\mathsf{D}_N$ and construct the SPM continued fraction interpolator
\begin{align}
	C_M(E)=\cfrac{I_1}{1+\cfrac{a_1(E-E_1)}{1+\cfrac{a_2(E-E_2)}{1+\cfrac{\cdots}{1+\cfrac{\cdots}{a_{M-1}(E-E_{M-1})}}}}},	
		\label{SPMcf}
\end{align}
where the $M-1$ coefficients $a_i$ are (recursively) constructed to ensure that \mbox{$C_M (E_i ) = I_i$}, $ \forall\ E_i\in \mathsf{D}_M$. The interpolator~\eqref{SPMcf} can be then cast in the form 
\begin{align}
	C_M(E)=\frac{P(E)}{Q(E)},
\end{align}
where $P$ and $Q$ are polynomials whose degree is determined by the size $M$ of the subset $\mathsf{D}_M$ chosen: $(M-1)/2$ ($P$ and $Q$) if M is odd; $M/2-1$ ($P$) and $M/2$ ($Q$) if $M$ is even. Thus, for large $E$,  $C_M\sim \mathrm{const.}$ or $C_M\sim 1/E$ respectively. Next, by simply letting $E$ take on complex values, $C_M(E)$ can be analytically continued to the complex plane; its pole structure can be easily determined, and will provide an approximation to the one of the original intensity $I(E)$.

We remark that the SPM and Pad\'e approximants share the same functional form: meromorphic functions possibly written as continued fractions. However, a Pad\'e approximant in its standard implementation is defined as an expansion of an analytic function near a specific point, \ie the coefficients are constructed from values of higher derivatives at a single point such that its power series agrees with the power series of the function it is approximating. Although the SPM will also result in a rational expression, the idea is different: the input is not the derivatives at a \emph{single} point, but the function values at \emph{different} points. Sometimes this is also referred to as a multipoint-Pad\'e method. In similar spirit, Ref.~\cite{Ropertz:2018stk} fits the coefficient of a Pad\'e functional form to a given theoretical model. In this way, the information from several data points is used, but statistical noise from the data is smoothened. In our approach, the resulting continued fraction interpolates through \emph{all} the points used for its construction. The SPM therefore has the advantage of being able to describe data points over a wide range of energy values without the need to compute any derivatives. The disadvantage is that a large number of spurious singularities is generated by the high order polynomial, which can be treated with the statistical procedure described in the following.

\begin{figure*}[!t]
	\includegraphics[width=2\columnwidth]{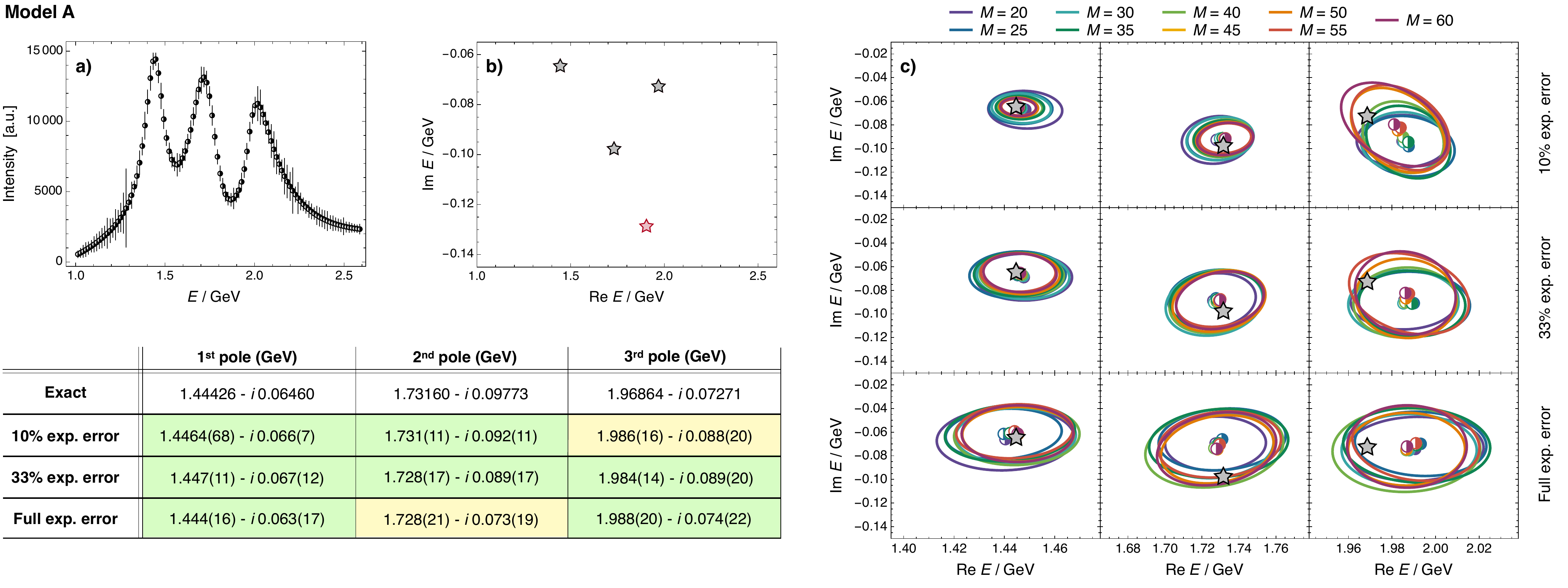}
	\caption{\label{fig:valid-model-A}({\it color online}) {\bf Validation of the statistical SPM method.} {\bf a)} The intensity dataset $\mathsf{D}_N$ generated by model A. {\bf b)} True pole structure of the intensity in panel a). The poles indicated in red here and in the figures below, represent artefacts of the validation models used, and are not associated to any physical resonance; with the exception of model C, they are correctly identified by the SPM only when exact data are used. {\bf c)} SPM determination of the data pole structure when considering $10\%$, $33\%$ and $100\%$ of the full experimental error (from top to bottom); half-filled circles correspond to the best value determinations of the pole position whereas ellipses indicate 68\% confidence regions  (defined as the isoprobability contour that  The different colors indicates the number of input points $M$ with \mbox{$M\in \{20+5i,i=0,1,\dots,8\}$}. Finally, in the table we show the SPM estimation of the poles' location (obtained as an average over the different $M$ which show a signal). The table color code indicates determinations where: both real and imaginary parts are compatible within  $1\sigma$  with the exact location (green); either the real or imaginary part are compatible within  $1\sigma$  with the exact location (yellow); neither one is compatible within $1\sigma$ with the exact location~(red). Notice that, with the exception of model B, table entries of the second and third type are always very close to the expected values, and in any case within $2\sigma$ from them.        
	}
\end{figure*}


How well SPM reconstructs the true structure mainly depends on the precision of the starting dataset $\mathsf{D}_N$,  and on the presence of other nonanalyticities such as branch points; for example, when exact numerical data are considered, $C_M$ gives rise to a pole structure which exactly reproduces\footnote{By construction, $Q(E)$ has additional poles with respect to the original $I(E)$. However, these have vanishing residue, as they are cancelled by corresponding zeroes of $P(E)$.} the one of $I$ independently from the number of input points $M$ (provided that the latter is large enough to capture the basic features of the intensity curve we wish to study). As for branch points, they seem not to affect the determination of poles sufficiently far from it. Interestingly, the SPM has been shown
to be able to detect also branch cuts,
in the form of a series of aligned poles~\cite{Binosi:2019ecz}.

In general, however, one is dealing with experimental datasets, where $\mathsf{D}_N$ comprises pairs that are only statistically distributed around the true intensity curve $I(E)$. The question is then whether or not the SPM is able to provide a veracious reconstruction of the corresponding pole structure even in such cases~\cite{Tripolt:2016cya, Chen:2018nsg, Binosi:2018rht, Binosi:2019ecz, Eichmann:2019dts, Yao:2020vef, Yao:2021pyf,Cui:2021vgm,Cui:2021aee,Cui:2021skn}, so that resonances can be determined in a purely data-driven approach which reduces to a minimum the theoretical bias introduced.

\begin{figure*}[!t]
	\includegraphics[width=2.09\columnwidth]{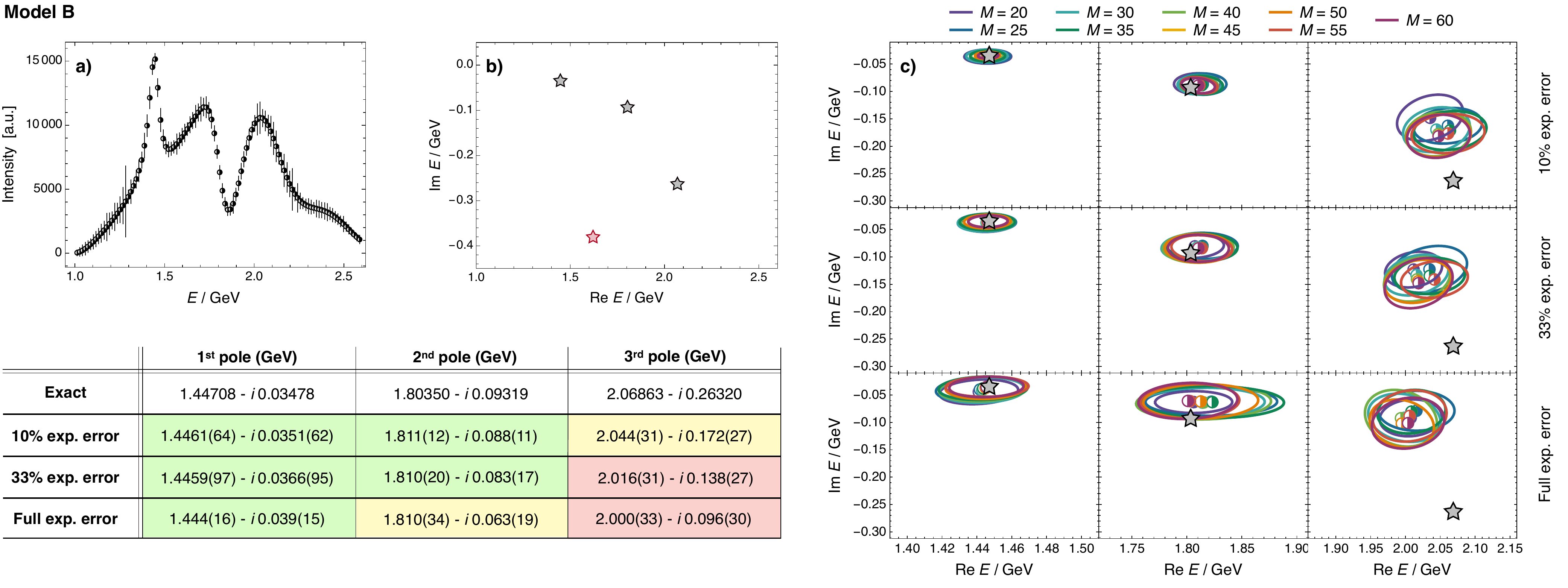}
	\caption{\label{fig:valid-model-B}({\it color online}) Same as in Fig.~\ref{fig:valid-model-A} but for model B. This turns out to be the most difficult model to for the SPM, which, in the case of the third pole, is incapable of providing reliable information on the imaginary part.             
	}
\end{figure*}

\begin{figure*}[!t]
	\includegraphics[width=2.09\columnwidth]{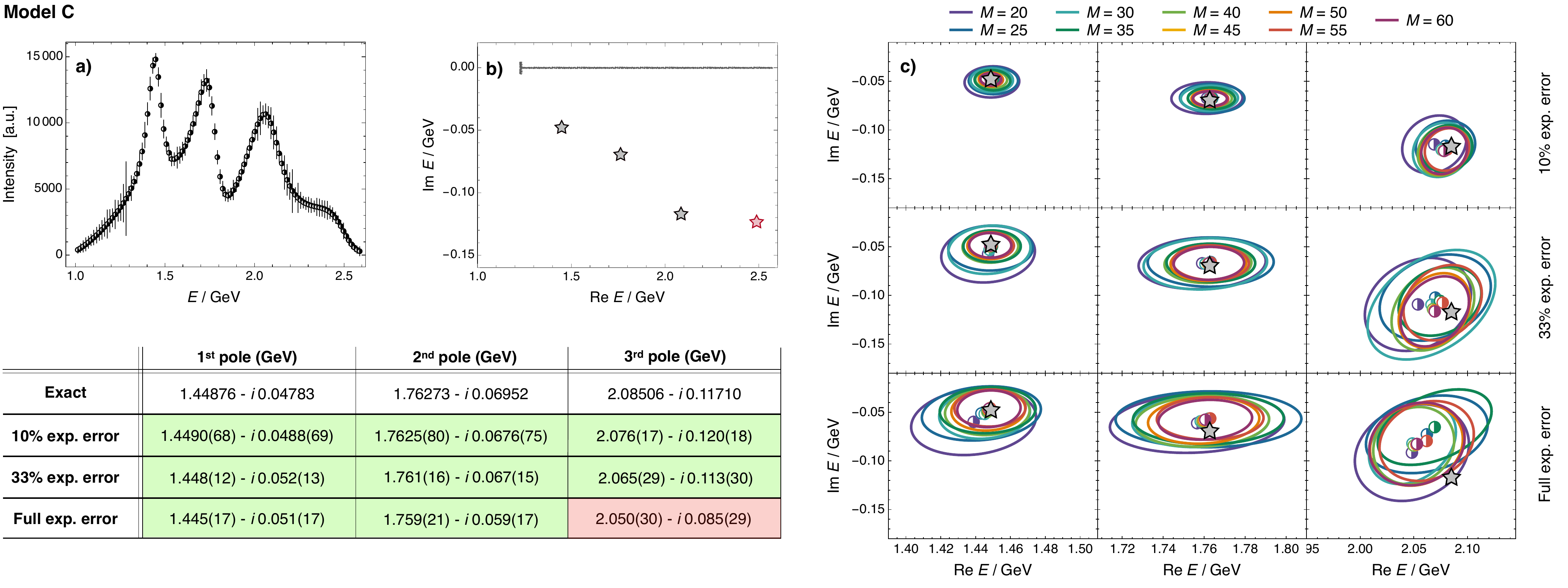}
	\caption{\label{fig:valid-model-C}({\it color online}) Same as in Fig.~\ref{fig:valid-model-A} but for model C. Notice the presence of the branch cut at the $\rho$ threshold, which however does not affect the SPM ability to find the pole locations. In this case, the SPM correctly determines also the position of the right-most pole (not shown in panel c) when considering 10\% and 33\% of the experimental error; at full experimental error, the signal for this pole fades away.       
	}
\end{figure*}

\begin{figure*}[!t]
	\includegraphics[width=2.09\columnwidth]{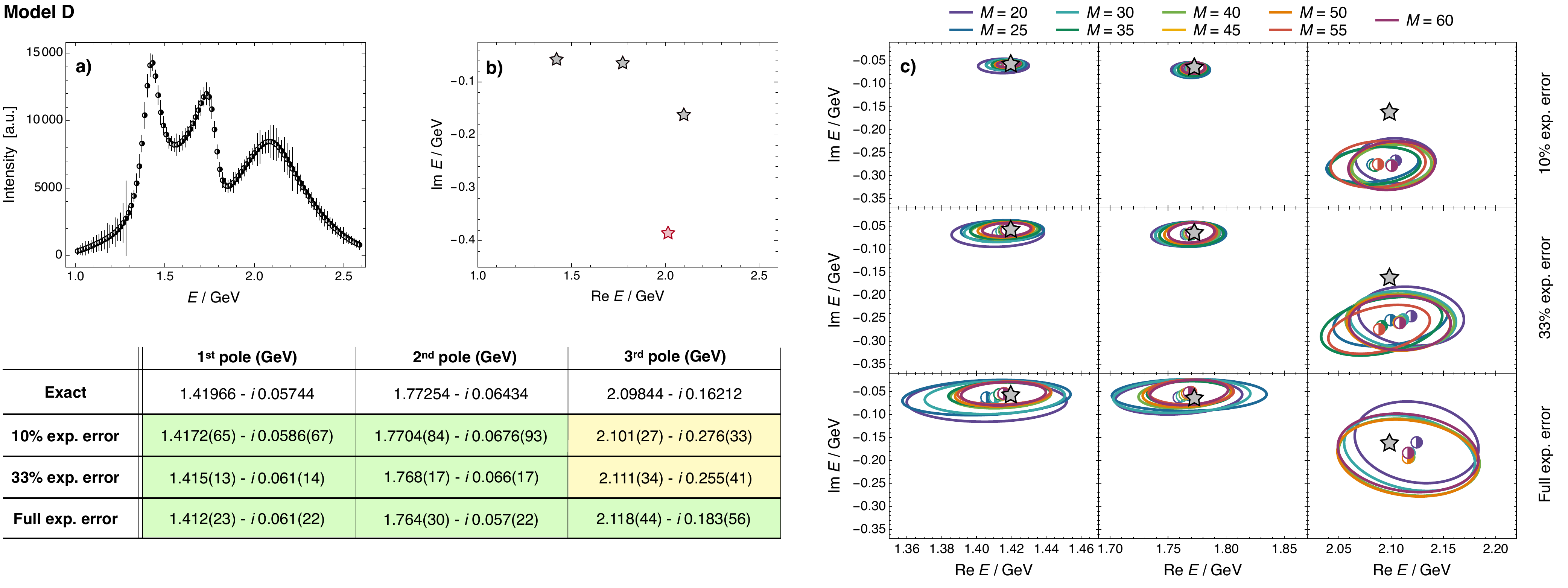}
	\caption{\label{fig:valid-model-D}({\it color online}) Same as in Fig.~\ref{fig:valid-model-A} but for model D. In this case the presence of the fourth pole (overlapping in mass with the third one) implies that for low error values the SPM returns a pole that is in between the two (panel c, top-right corner); as the error increases the signal of the fourth pole faints and the correct position of the pole is recovered.  
	}
\end{figure*}

\begin{figure*}[!t]
	\includegraphics[width=2.09\columnwidth]{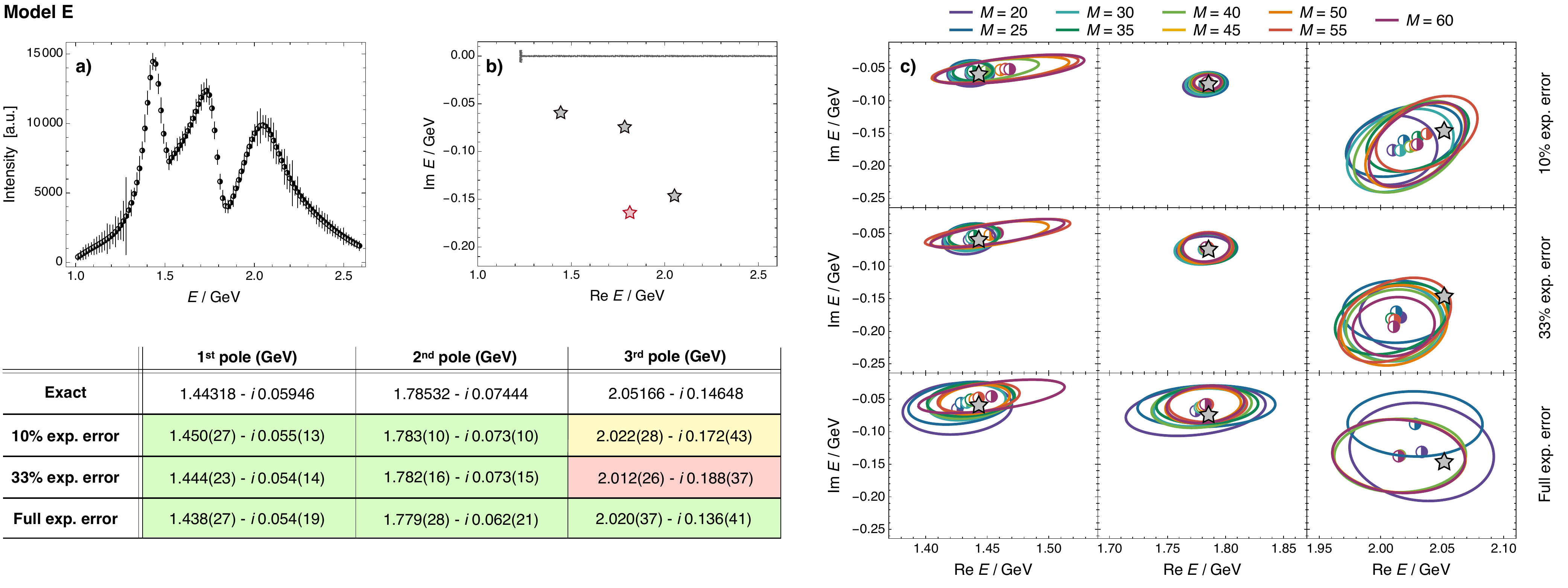}
	\caption{\label{fig:valid-model-E}({\it color online}) Same as in Fig.~\ref{fig:valid-model-A} but for model E. As was already the case for model C, this model features a cut at the $\rho$ threshold, which (again) does not prevent the SPM to reconstruct its pole structure. For the first pole at the largest $M$ a not fully resolved structure appear, causing the oblation of the signal region. For the third pole and full experimental error, only even values of $M$ ({\it i.e.}, the ones with a vanishing asymptotic) resolve the pole.          
	}
\end{figure*}

To answer this question, we consider the combination of the SPM with a bootstrap procedure~\cite{Cui:2021vgm}. Let us consider the set of pairs $\mathsf{D}_N$, with associated statistical error (for example see Fig.~\ref{fig:proc}a). From this we can generate a replica set  $\mathsf{D}^r_N$ by randomly drawing from each datum in $\mathsf{D}_N$ a new one normally distributed around a mean value equal to the datum itself and a standard deviation equal to its associated error (Fig.~\ref{fig:proc}b). At this point we randomly choose the subset $\mathsf{D}^r_M\subseteq\mathsf{D}^r_N$ (Fig.~\ref{fig:proc}c) and proceed to construct the corresponding SPM interpolator $C_M^r$ (Fig.~\ref{fig:proc}d). We can next determine all the poles of  $C_M^r$ from the zeroes of $Q(E)$; since the coefficients $a_i$ of the interpolator are real, these poles appear in complex conjugate pairs,\footnote{We remark that the conjugate pairs mentioned here are not the ones of the amplitude generated by Schwartz reflection principle, but rather the ones appearing in its modulus squared, \ie the intensity, as shown in Eq.~\eqref{eq:intensity}.} of which it is sufficient to consider the ones in the lower half of the complex plane only (Fig.~\ref{fig:proc}e).

Repeating steps b) through d) for a sufficiently large number of replicas $n_r$ (we use $n_r=10^5$ in the ensuing analysis), we pass from the single pole snapshot of Fig.~\ref{fig:proc}e to the distribution shown in Fig.~\ref{fig:diffimg}a which comprises $\sim5.5\times10^5$ poles. 
In fact, statistical fluctuations imply that the SPM interpolating fractions have additional wiggles/jumps (see again Fig.~\ref{fig:proc}d) that correspond to additional {\it noise-poles} that are absent for exact data. At this point the question is then how to filter out the {\it signal poles} from these latter ones. 

To this purpose, we first filter the result to isolate poles' clusters; this can be simply achieved by applying transparency to each point representing an SPM identified pole, which efficiently highlights denser poles regions (Fig.~\ref{fig:diffimg}a and~b).
Next, we apply morphological pattern recognition to extract image components representing the boundaries of the identified regions:\footnote{This is achieved through {\tt Mathematica} by: {\it i}) {\tt ColorNegate} a binarized image of the pole clusters obtained through the {\tt MorphologicalBinarize} command; {\it ii}) obtaining an array in which each pixel of this binarized image is replaced by an integer index representing the connected foreground image component in which the pixel lies (through the {\tt MorphologicalComponent} command); and {\it iii}) computing the ``Centroid'', ``SemiAxes'', and ``Orientation'' properties for the image components using the {\tt ComponentMeasurements} command.} their centroid, the length of the semi-axes of the best fitting ellipse and its orientation (defined as the angle between the largest axis and the horizontal axis). There are cases in which noise poles in general, and the ones that accumulate near the real axis in particular, are mistakenly grouped in clusters of various sizes (Fig.~\ref{fig:diffimg}c). However, such noise poles clusters have an orientation which is very close to zero (they are disposed almost horizontally, whereas signal pole are almost vertically aligned) and/or a very small size, so that they can be easily identified and removed.

Poles that lie within the geometric region identified  (Fig.~\ref{fig:diffimg}d) are then considered to be the ones that are approximating the true poles present in the analytic structure of the intensity we are studying. The last step consists in selecting those poles and construct through the mean value and covariance matrix of each cluster the standard ellipsoids for a given confidence level.\footnote{The ellipse is the isoprobability contour of a multivariate gaussian distribution having for mean value and covariance matrix the ones obtained with the standard estimators. We define confidence level the probability enclosed by this contour. This implies that, for this 2D case, the semiaxes of a $1\sigma$ ellipse are $\approx 1.52$ times the square root of the eigenvalues of the covariance matrix.}

Finally, we will not specify the number of SPM input points $M$, studying instead how the SPM results are (in)dependent on its value.

\begin{figure*}[!t]
	\includegraphics[width=2.1\columnwidth]{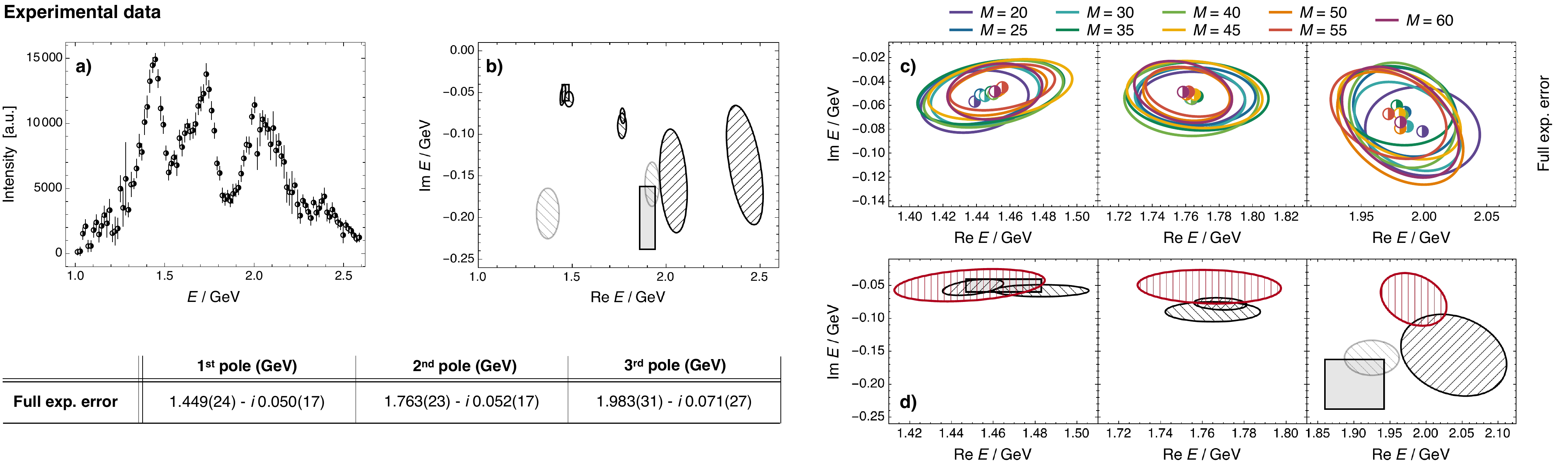}
	\caption{\label{fig:exp-extraction}({\it color online}) {\bf SPM extraction of resonance poles from BESIII data.} {\bf a)} The intensity dataset $\mathsf{D}_N$ obtained from the $J/\psi\to\gamma\pi^0\pi^0$ radiative decays data of BESIII~\cite{BESIII:2015rug}. {\bf b)} $1\sigma$ confidence level regions for the positions of the first three scalar resonances obtained from various analyses (see also Table~\ref{tab:mandw}): JPAC~\cite{Rodas:2021tyb} ($\pi/4$ hatchings); and Bonn-Gatchina~\cite{Sarantsev:2021ein} including two possible candidates for $f_0(1500)$ and $f_0(2020)$ ($-\pi/4$ hatchings). We also plot the determination of $f_0(1500)$ and $f_0(2020)$ by Ropertz {\it et al.}~\cite{Ropertz:2018stk} (filled rectangles) obtained from analysis of the $B_s \to J/\psi \pi \pi$ decay. In the case of the Bonn-Gatchina, the contours are built assuming uncorrelated mass and width, whereas for the Ropertz {\it et al.}, the rectangles represent the envelope of the $1\sigma$ confidence level elliptical regions obtained for the pole position from the multiple fits performed (see~\cite{Ropertz:2018stk} for details). {\bf c)} SPM determination of the pole structure for different $M$ (all ellipses indicating the $1\sigma$ confidence regions as usual). The SPM estimation of the poles' location (obtained as an average over the different $M$ which provide a signal) is reported in the Table on the left. {\bf d)}~Comparison of the SPM $1\sigma$ confidence level  regions  for the determinations of the $f_0$ scalar resonances (red, vertical hatching) with the different determination shown in panel b).        
	}
\end{figure*}

\section{\label{sec:validation}Validation}  

The statistical SPM procedure described above can be validated by applying it to data generated from five different models (labelled from A to E) \cite{Rodas:2021tyb} that were developed to perform a combined analysis of the scalar and tensor waves of $J/\psi \to \gamma \pi^0 \pi^0$ and $\to \gamma K_S^0 K_S^0$~\cite{BESIII:2015rug,BESIII:2018ubj}. They contain either three or four poles in the complex energy plane, with two of them (C and E)  including also a stable $\rho\rho$ channel, which gives a branch point on the real axis at $E = 1.52\gev$. They are defined as follows.

The model's amplitudes are given by
\begin{equation}
\label{eq:amplitude}
 a^J_i(E) = E_\gamma\, p_i^{J} \, \sum_k n^J_k(E) \left[ {D^J(E)}^{-1} \right]_{ki},
\end{equation}
with the index $i=h\bar h=\pi\pi$, $K\bar K$, and possibly $\rho\rho$ for models B and E; as customary, $E$ is the $h\bar h$ invariant mass,
$p_i=\sqrt{E^2- 4m_{i}^2}/2$ and
$E_\gamma=(m_{\jpsi}^2-E^2)/(2E)$. We use $m_\pi = 134.98\mev$, $m_K = 497.61\mev$, $m_\rho=762\mev$.
The numerator $n^J_k(E)$ is effectively expanded as
\begin{equation}
    n^J_k(E)=\sum_{n=0}^{3}{a^{J,k}_n  T_n\left[\omega(E)\right]},
\label{eq:production}
\end{equation}
where $T_n$ are the Chebyshev polynomials of order $n$. The models have different choices of $\omega(E)$,
\begin{subequations}
\begin{align}
\omega(E)_\text{model A,B,C}&=2\frac{E^2-E^2_\text{min}}{E^2_\text{max}-E^2_\text{min}}-1\,, \label{eq:omegascaled}\\
\omega(E)_\text{model D,E} &=2\frac{\omega(E)_\text{pole}-\omega(E_\text{min})_\text{pole}}{\omega(E_\text{min})_\text{pole}-\omega(E_\text{max})_\text{pole}}-1, \label{eq:omegapolescaled}
\end{align}
\end{subequations}
where $E_0 = E_\text{min} = 1\gev$, $E_\text{max} = 2.5\gev$, and $\omega(E)_\text{pole} = E^2/(E^2 + E_0^2)$. The denominator is given by
\begin{equation}
\label{eq:Dsol}
D^J_{ki}(E) =  \left[ {K^J(E)}^{-1}\right]_{ki} - \frac{E^2}{\pi}\int_{4m_{k}^2}^{\infty}ds'\frac{\rho N^J_{ki}(s') }{s'(s'-E^2 - i\epsilon)}, 
\end{equation}
with
\begin{align}
\rho N^J_{ki}(s')  &= \delta_{ki} \,\frac{(2p_i)^{2J+1}}{\left(s'+s_L\right)^{2J+1}},
\end{align}
where $J=0$ for the scalar wave and for the $\rho\rho$ channel for the tensor wave, and $J=2$ for the other channel in the tensor wave. For model A and B $s_L = 0$ , whereas $s_L = 0.6\gev^2$ for models C, D, E.
For the $K$-matrix, we consider 
\begin{subequations}
\begin{align}
K^J_{ki}(E)_\text{model B,D,E} &= \sum_R \frac{g^{J,R}_k g^{J,R}_i}{m_R^2 - E^2} + c^J_{ki} + d^J_{ki} \,E^2,\label{eq:Kmatrix}
\end{align}
with $c^J_{ki}= c^J_{ik}$ and $d^J_{ki}= d^J_{ik}$.
Alternatively,the inverse of the $S$-wave $K$-matrix is parametrized as a sum of CDD poles~\cite{Castillejo:1955ed,JPAC:2017dbi},
\begin{equation}
\left[K^J(E)^{-1}\right]_{ki}^\text{model A,C} = c^J_{ki} - d^J_{ki} \,E^2 - \sum_R \frac{g^{J,R}_k g^{J,R}_i}{m_R^2 - E^2}, \label{eq:CDD}
\end{equation}
\end{subequations}
where $c^J_{ki} = c^J_{ik}$ and $d^J_{ki} = d^J_{ik}$ are constrained to be positive. All needed models' parameters are reported in Table~\ref{tab:values1} of Appendix~\ref{app:parameters}. For each analytic model, we calculate the value of the intensity in correspondence of the experimental energy bin. To study the SPM performance in presence of realistic statistical uncertainties, we associate to each data point 10\%, 33\% or 100\% of the experimental error; the method is finally applied to data generated in the energy region $1.0 \le E \le 2.5\gev$, which is within the range where the models of~\cite{Rodas:2021tyb} were fitted to the experimental results.

\begin{table*}[!t]
\renewcommand\arraystretch{1.3}
\begin{tabular}{|c|c|c|c|c|}
 \hline
     & SPM (this work)    & JPAC    & Bonn-Gatchina     & Ropertz {\it et al.}    \\
 \hline
 $f_0(1500)$         & 
 $(1449\pm24)-i(100\pm 32)/2$    & $(1450\pm10)-i(106\pm 16)/2$    & $(1483\pm15)-i(116\pm 12)/2$   & $(1465\pm18)-i(101\pm20)/2$    \\
 \hline
 $f_0(1710)$         & $(1763\pm 23)-i(104\pm 34)/2$    & $(1769\pm8)-i(156\pm 12)/2$    & $(1765\pm15)-i(180\pm 20)/2$    & /    \\
 \hline
 \multirow{2}*{$f_0(2020)$ }         & \multirow{2}*{$(1983\pm 31)-i(143\pm54)/2$}    & \multirow{2}*{$(2038\pm48)-i(312\pm 82)/2$}    & $(1925\pm25)-i(320\pm 35)/2$   & \multirow{2}*{$(1901\pm41)-i(401\pm76)/2$}    \\
 & & & $(2075\pm20)-i(260\pm 25)/2$ & \\
\hline
\end{tabular}
\caption{\label{tab:mandw}Pole positions for the first three scalar resonances obtained from various data analyses (all results expressed in MeV). In addition to the SPM  and JPAC~\cite{Rodas:2021tyb} determinations, we show the Bonn-Gatchina~\cite{Sarantsev:2021ein} (with two possible candidates for $f_0(1500)$ and $f_0(2020)$) and Ropertz {\it et al.}~\cite{Ropertz:2018stk} ones. 
}
\end{table*}


As shown in Figs.~\ref{fig:valid-model-A} through~\ref{fig:valid-model-E}, the SPM is always able to locate the position of the first two poles with precision and accuracy; the third pole is more difficult; however,  the   SPM reconstructed location is in all but one case well within $2\sigma$ from the true value. Also, as observed in other studies~\cite{Tripolt:2016cya,Tripolt:2018xeo,Chen:2018nsg,Binosi:2018rht, Binosi:2019ecz,Eichmann:2019dts,Yao:2020vef, Yao:2021pyf,Cui:2021vgm,Cui:2021aee,Cui:2021skn}, the SPM results are by and large independent from the value of input points $M$, with both the central values and the  $1\sigma$ confidence level contours showing very little variation with $M$ for all validation models. 

Overall, the SPM algorithm devised can be considered to be validated by the study presented: one obtains excellent results for the first two poles for all models considered, and a very good determination of the third one. Clearly, the size of the experimental uncertainties is the key limiting factor for a precise pole structure determination through the SPM. 

\section{Pole determination from experimental data}
\label{sec:results}

When applying the SPM procedure to the $J/\psi\to\gamma\pi^0\pi^0$ radiative decay data of BESIII~\cite{BESIII:2015rug}, we obtain the results shown in Fig.~\ref{fig:exp-extraction}. The SPM detects in this case three poles at the coordinates listed therein, presumably connected to the $f_0(1500)$, $f_0(1710)$ and $f_0(2020)$ states. The mass and width values with the associated  uncertainties are also reported in Tab.~\ref{tab:mandw}, together with other (model-dependent) recent determinations~\cite{Rodas:2021tyb,Sarantsev:2021ein,Ropertz:2018stk}. As can be seen the SPM results compare well with the latter, especially the JPAC values; however, within the current data precision, there seems to be a tendency by the SPM to provide smaller widths for the higher mass resonances. We remark that the other determinations include more information from other channels, while here we  consider the $J/\psi\to\gamma\pi^0\pi^0$ data only.
A graphical comparison of the results is provided in Fig.~\ref{fig:exp-extraction}d.

\section{Summary}
\label{sec:summary}

We presented a study of a data-driven method for the determination of complex poles associated to resonances from experimental data. The algorithm is based on the Schlessinger Point Method~\cite{PhysRev.167.1411, Schlessinger:1966zz}, which interpolates a given dataset by the continued-fraction expression~(\ref{SPMcf}), for then evaluating the resulting function at complex arguments to gain access to its pole structure~\cite{Tripolt:2016cya,Tripolt:2018xeo,Chen:2018nsg,Binosi:2018rht, Binosi:2019ecz,Eichmann:2019dts,Yao:2020vef, Yao:2021pyf,Cui:2021vgm,Cui:2021aee,Cui:2021skn}.

The method, once specifically adapted to deal with reactions of interest for spectroscopy, has been benchmarked against the available experimental data on the $J/\psi \to \gamma \pi^0\pi^0$ decay measured by BESIII~\cite{BESIII:2015rug}, which, within the investigated energy range, are expected to describe three scalar resonances: $f_0(1500)$, $f_0(1710)$, and $f_0(2020)$. Validation has been first obtained via reproducing the analytical structure of the pseudo data generated from five different models that were proposed to describe the BESIII results~\cite{Rodas:2021tyb}. 

The results obtained demonstrate the SPM's capability to reliably identify the presence and characterize the mass and width of resonances in reaction amplitudes already at the current experimental precision level. In prevision of higher statistics datasets, our study shows that the SPM will compete in precision and accuracy with traditional (model dependent) spectroscopy analysis techniques, complementing them towards a robust determination and characterization of resonances. 
The method has been applied to
single channel intensities so far, and the future challenge
is to extend it to study coupled-channel datasets and, possibly, the
relative phase of different partial waves.

\begin{acknowledgments}
 
We thank Arkaitz Rodas for useful exchanges and one of the anonymous reviewer for useful suggestions on how to improve the analysis presented. R.-A.~T.~is supported by the Deutsche Forschungsgemeinschaft (DFG, German Research Foundation) through the Collaborative Research Center CRC-TR 211 ``Strong-interaction matter under extreme conditions'' -- Project No. 315477589-TRR 211. 
 
\end{acknowledgments}
 
\appendix 
\section{Model parameters}
\label{app:parameters}

In this Appendix we list explicit numerical values used for the five models discussed in Sec.~\ref{sec:validation} which fit the experimental data on the $J/\psi\to\gamma \pi^0\pi^0$ radiative decay measured by BESIII~\cite{BESIII:2015rug} and which were proposed in \cite{Rodas:2021tyb}. The various parameters are listed in Tab.~\ref{tab:values1} and Tab.~\ref{tab:values2}, see Sec.~\ref{sec:validation} and \cite{Rodas:2021tyb} for details.

\begin{table*}[h]
\begin{ruledtabular}
\begin{tabular}{c c c c c c}
 & A & B & C & D & E\\ \hline
\multicolumn{6}{c}{$\pi \pi$ channel}  \\ \hline
\rule[-0.2cm]{-0.1cm}{.55cm} $a^{S,\pi\pi}_0$ & $7.231$ & $-15.666$ & $157.245$ & $-1227.005$ & $-2173.729$ \\
\rule[-0.2cm]{-0.1cm}{.55cm} $a^{S,\pi\pi}_1$ & $99.992$ & $303.153$ & $162.476$ & $1574.019$ & $3272.052$ \\
\rule[-0.2cm]{-0.1cm}{.55cm} $a^{S,\pi\pi}_2$ & $64.425$ & $63.199$ & $62.922$ & $-848.475$ & $-1553.726$ \\
\rule[-0.2cm]{-0.1cm}{.55cm} $a^{S,\pi\pi}_3$ & $-25.603$ & $53.894$ & $50.439$ & $58.143$ & $361.790$ \\
\rule[-0.2cm]{-0.1cm}{.55cm} $a^{D,\pi\pi}_0$ & $4.320$ & $62.803$ & $-604.559$ & $97.022$ & $109.759$ \\
\rule[-0.2cm]{-0.1cm}{.55cm} $a^{D,\pi\pi}_1$ & $-247.631$ & $-48.166$ & $-816.649$ & $-11.778$ & $-2.844$ \\
\rule[-0.2cm]{-0.1cm}{.55cm} $a^{D,\pi\pi}_2$ & $-62.318$ & $-22.265$ & $-276.267$ & $61.621$ & $71.095$ \\
\rule[-0.2cm]{-0.1cm}{.55cm} $a^{D,\pi\pi}_3$ & $-99.560$ & $20.785$ & $-47.735$ & $0$ & $0$ \\
\hline 
\multicolumn{6}{c}{$K \bar K$ channel}  \\ \hline
\rule[-0.2cm]{-0.1cm}{.55cm} $a^{S,K\bar K}_0$ & $110.543$ & $-59.382$ & $150.521$ & $1308.089$ & $1927.495$ \\
\rule[-0.2cm]{-0.1cm}{.55cm} $a^{S,K\bar K}_1$ & $412.680$ & $258.425$ & $118.193$ & $-1875.100$ & $-2996.838$ \\
\rule[-0.2cm]{-0.1cm}{.55cm} $a^{S,K\bar K}_2$ & $144.309$ & $-80.752$ & $54.494$ & $827.515$ & $1354.327$ \\
\rule[-0.2cm]{-0.1cm}{.55cm} $a^{S,K\bar K}_3$ & $-128.449$ & $33.021$ & $70.733$ & $-53.201$ & $-287.158$ \\
\rule[-0.2cm]{-0.1cm}{.55cm} $a^{D,K\bar K}_0$ & $-33.051$ & $28.522$ & $309.966$ & $-28.364$ & $-26.964$ \\
\rule[-0.2cm]{-0.1cm}{.55cm} $a^{D,K\bar K}_1$ & $-76.768$ & $-12.414$ & $523.716$ & $-92.341$ & $-135.324$ \\
\rule[-0.2cm]{-0.1cm}{.55cm} $a^{D,K\bar K}_2$ & $-153.380$ & $-7.750$ & $294.291$ & $-16.016$ & $-14.887$ \\
\rule[-0.2cm]{-0.1cm}{.55cm} $a^{D,K\bar K}_3$ & $-35.367$ & $5.898$ & $74.811$ & $0$ & $0$ \\
\hline 
\multicolumn{6}{c}{$S$-wave couplings} \\ \hline
\rule[-0.2cm]{-0.1cm}{.55cm} $g^{S,1}_{\pi\pi}$ & $-0.233$ & $0.172$ & $0.651$ & $-6.398$ & $-6.058$ \\
\rule[-0.2cm]{-0.1cm}{.55cm} $g^{S,1}_{K\bar K}$ & $0.388$ & $-0.063$ & $0.931$ & $-6.329$ & $-7.247$ \\
\rule[-0.2cm]{-0.1cm}{.55cm} $g^{S,1}_{\rho\rho}$ & $0$ & $0$ & $0.610$ & $0$ & $1.956$ \\
\rule[-0.2cm]{-0.1cm}{.55cm} $g^{S,2}_{\pi\pi}$ & $0.588$ & $0.305$ & $0.223$ & $-0.235$ & $-0.236$ \\
\rule[-0.2cm]{-0.1cm}{.55cm} $g^{S,2}_{K\bar K}$ & $0.345$ & $0.681$ & $-0.188$ & $0.657$ & $-1.092$ \\
\rule[-0.2cm]{-0.1cm}{.55cm} $g^{S,2}_{\rho\rho}$ & $0$ & $0$ & $-0.314$ & $0$ & $-0.233$ \\
\rule[-0.2cm]{-0.1cm}{.55cm} $g^{S,3}_{\pi\pi}$ & $1.179$ & $0.402$ & $1.219$ & $9.166$ & $8.295$ \\
\rule[-0.2cm]{-0.1cm}{.55cm} $g^{S,3}_{K\bar K}$ & $0.910$ & $-0.755$ & $1.263$ & $13.967$ & $12.908$ \\
\rule[-0.2cm]{-0.1cm}{.55cm} $g^{S,3}_{\rho\rho}$ & $0$ & $0$ & $-0.292$ & $0$ & $-1.049$ \\
\hline 
\multicolumn{6}{c}{$D$-wave couplings} \\ \hline
\rule[-0.2cm]{-0.1cm}{.55cm} $g^{D,1}_{\pi\pi}$ & $0.482$ & $-0.995$ & $2.381$ & $0.944$ & $0.155$ \\
\rule[-0.2cm]{-0.1cm}{.55cm} $g^{D,1}_{K\bar K}$ & $0.275$ & $-0.391$ & $6.949$ & $0.979$ & $1.027$ \\
\rule[-0.2cm]{-0.1cm}{.55cm} $g^{D,1}_{\rho\rho}$ & $0$ & $0$ & $61.196$ & $0$ & $-0.344$ \\
\rule[-0.2cm]{-0.1cm}{.55cm} $g^{D,2}_{\pi\pi}$ & $-0.085$ & $-9.881$ & $0.137$ & $-0.897$ & $0.908$ \\
\rule[-0.2cm]{-0.1cm}{.55cm} $g^{D,2}_{K\bar K}$ & $0.831$ & $27.650$ & $1.463$ & $0.310$ & $-0.250$ \\
\rule[-0.2cm]{-0.1cm}{.55cm} $g^{D,2}_{\rho\rho}$ & $0$ & $0$ & $-5.565$ & $0$ & $0$ \\
\rule[-0.2cm]{-0.1cm}{.55cm} $g^{D,3}_{\pi\pi}$ & $0.516$ & $0.063$ & $1.061$ & $-0.009$ & $0.852$ \\
\rule[-0.2cm]{-0.1cm}{.55cm} $g^{D,3}_{K\bar K}$ & $-0.345$ & $-0.943$ & $0.350$ & $0.974$ & $0.698$ \\
\rule[-0.2cm]{-0.1cm}{.55cm} $g^{D,3}_{\rho\rho}$ & $0$ & $0$ & $0$ & $0$ & $1.134$ \\
\end{tabular}
\end{ruledtabular}
\caption{\label{tab:values1} Numerical values for the parameters used for the models discussed in Sec.~\ref{sec:validation} which fit the experimental data on the $J/\psi\to\gamma \pi^0\pi^0$ radiative decay measured by BESIII~\cite{BESIII:2015rug} and which were proposed in \cite{Rodas:2021tyb}.}
\end{table*}

\begin{table*}[h]
\begin{ruledtabular}
\begin{tabular}{c c c c c c}
\multicolumn{6}{c}{$S$-wave bare masses}   \\ \hline
\rule[-0.2cm]{-0.1cm}{.55cm} $m^2_{S,1}$ & $2.068$ & $2.087$ & $4.247$ & $0.036$ & $0.006$ \\
\rule[-0.2cm]{-0.1cm}{.55cm} $m^2_{S,2}$ & $3.102$ & $3.012$ & $2.293$ & $3.120$ & $3.184$ \\
\rule[-0.2cm]{-0.1cm}{.55cm} $m^2_{S,3}$ & $4.893$ & $3.762$ & $6.406$ & $8.239$ & $7.548$ \\
\hline
\multicolumn{6}{c}{$D$-wave bare masses}   \\ \hline
\rule[-0.2cm]{-0.1cm}{.55cm} $m^2_{D,1}$ & $1.392$ & $1.080$ & $19.512$ & $2.848$ & $2.330$ \\
\rule[-0.2cm]{-0.1cm}{.55cm} $m^2_{D,2}$ & $2.285$ & $15.775$ & $2.423$ & $1.674$ & $1.667$ \\
\rule[-0.2cm]{-0.1cm}{.55cm} $m^2_{D,3}$ & $3.833$ & $2.322$ & $1.498$ & $2.324$ & $2.894$ \\
\hline
\multicolumn{6}{c}{$S$-wave background} \\ \hline
\rule[-0.2cm]{-0.1cm}{.55cm} $c^{S,0}_{\pi\pi,\pi\pi}$ & $-0.001$ & $-0.870$ & $1.081$ & $21.683$ & $16.190$ \\
\rule[-0.2cm]{-0.1cm}{.55cm} $c^{S,0}_{\pi\pi,K\bar K}$ & $2.088$ & $0.468$ & $1.820$ & $18.757$ & $18.308$ \\
\rule[-0.2cm]{-0.1cm}{.55cm} $c^{S,0}_{\pi\pi,\rho\rho}$ & $0$ & $0$ & $0$ & $0$ & $0$ \\
\rule[-0.2cm]{-0.1cm}{.55cm} $c^{S,0}_{K\bar K,K\bar K}$ & $-3.745$ & $-0.558$ & $2.063$ & $12.777$ & $19.239$ \\
\rule[-0.2cm]{-0.1cm}{.55cm} $c^{S,0}_{K\bar K,\rho\rho}$ & $0$ & $0$ & $0$ & $0$ & $0$ \\
\rule[-0.2cm]{-0.1cm}{.55cm} $c^{S,0}_{\rho\rho,\rho\rho}$ & $0$ & $0$ & $1.185$ & $0$ & $-15.811$ \\
\rule[-0.2cm]{-0.1cm}{.55cm} $c^{S,1}_{\pi\pi,\pi\pi}$ & $0.425$ & $0.119$ & $0.381$ & $-7.554$ & $-6.274$ \\
\rule[-0.2cm]{-0.1cm}{.55cm} $c^{S,1}_{\pi\pi,K\bar K}$ & $0.939$ & $-0.014$ & $0.784$ & $-9.592$ & $-9.197$ \\
\rule[-0.2cm]{-0.1cm}{.55cm} $c^{S,1}_{\pi\pi,\rho\rho}$ & $0$ & $0$ & $0$ & $0$ & $0$ \\
\rule[-0.2cm]{-0.1cm}{.55cm} $c^{S,1}_{K\bar K,K\bar K}$ & $1.959$ & $-0.162$ & $-0.887$ & $-12.256$ & $-13.371$ \\
\rule[-0.2cm]{-0.1cm}{.55cm} $c^{S,1}_{K\bar K,\rho\rho}$ & $0$ & $0$ & $0$ & $0$ & $0$ \\
\rule[-0.2cm]{-0.1cm}{.55cm} $c^{S,1}_{\rho\rho,\rho\rho}$ & $0$ & $0$ & $0.534$ & $0$ & $6.031$ \\
\hline
\multicolumn{6}{c}{$D$-wave background} \\ \hline
\rule[-0.2cm]{-0.1cm}{.55cm} $c^{D,0}_{\pi\pi,\pi\pi}$ & $-0.286$ & $-4.408$ & $2.643$ & $-1.356$ & $-1.170$ \\
\rule[-0.2cm]{-0.1cm}{.55cm} $c^{D,0}_{\pi\pi,K\bar K}$ & $0.043$ & $16.881$ & $-0.723$ & $-0.960$ & $-1.012$ \\
\rule[-0.2cm]{-0.1cm}{.55cm} $c^{D,0}_{\pi\pi,\rho\rho}$ & $0$ & $0$ & $0$ & $0$ & $0$ \\
\rule[-0.2cm]{-0.1cm}{.55cm} $c^{D,0}_{K\bar K,K\bar K}$ & $0.921$ & $-50.498$ & $-1.287$ & $1.440$ & $0.395$ \\
\rule[-0.2cm]{-0.1cm}{.55cm} $c^{D,0}_{K\bar K,\rho\rho}$ & $0$ & $0$ & $0$ & $0$ & $0$ \\
\rule[-0.2cm]{-0.1cm}{.55cm} $c^{D,0}_{\rho\rho,\rho\rho}$ & $0$ & $0$ & $-820.930$ & $0$ & $-11.365$ \\
\rule[-0.2cm]{-0.1cm}{.55cm} $c^{D,1}_{\pi\pi,\pi\pi}$ & $-0.010$ & $-0.983$ & $-0.878$ & $0.223$ & $0.211$ \\
\rule[-0.2cm]{-0.1cm}{.55cm} $c^{D,1}_{\pi\pi,K\bar K}$ & $0.086$ & $1.631$ & $-0.558$ & $0.141$ & $0.168$ \\
\rule[-0.2cm]{-0.1cm}{.55cm} $c^{D,1}_{\pi\pi,\rho\rho}$ & $0$ & $0$ & $0$ & $0$ & $0$ \\
\rule[-0.2cm]{-0.1cm}{.55cm} $c^{D,1}_{K\bar K,K\bar K}$ & $-0.246$ & $-3.355$ & $-0.629$ & $-0.179$ & $0.005$ \\
\rule[-0.2cm]{-0.1cm}{.55cm} $c^{D,1}_{K\bar K,\rho\rho}$ & $0$ & $0$ & $0$ & $0$ & $0$ \\
\rule[-0.2cm]{-0.1cm}{.55cm} $c^{D,1}_{\rho\rho,\rho\rho}$ & $0$ & $0$ & $128.239$ & $0$ & $3.923$ \\
\end{tabular}
\end{ruledtabular}
\caption{\label{tab:values2} Numerical values for the parameters used for the models discussed in Sec.~\ref{sec:validation} which fit the experimental data on the $J/\psi\to\gamma\pi^0\pi^0$ radiative decay measured by BESIII~\cite{BESIII:2015rug} and which were proposed in \cite{Rodas:2021tyb}.}
\end{table*}



\begin{thebibliography}{46}%
\makeatletter
\providecommand \@ifxundefined [1]{%
 \@ifx{#1\undefined}
}%
\providecommand \@ifnum [1]{%
 \ifnum #1\expandafter \@firstoftwo
 \else \expandafter \@secondoftwo
 \fi
}%
\providecommand \@ifx [1]{%
 \ifx #1\expandafter \@firstoftwo
 \else \expandafter \@secondoftwo
 \fi
}%
\providecommand \natexlab [1]{#1}%
\providecommand \enquote  [1]{``#1''}%
\providecommand \bibnamefont  [1]{#1}%
\providecommand \bibfnamefont [1]{#1}%
\providecommand \citenamefont [1]{#1}%
\providecommand \href@noop [0]{\@secondoftwo}%
\providecommand \href [0]{\begingroup \@sanitize@url \@href}%
\providecommand \@href[1]{\@@startlink{#1}\@@href}%
\providecommand \@@href[1]{\endgroup#1\@@endlink}%
\providecommand \@sanitize@url [0]{\catcode `\\12\catcode `\$12\catcode
  `\&12\catcode `\#12\catcode `\^12\catcode `\_12\catcode `\%12\relax}%
\providecommand \@@startlink[1]{}%
\providecommand \@@endlink[0]{}%
\providecommand \url  [0]{\begingroup\@sanitize@url \@url }%
\providecommand \@url [1]{\endgroup\@href {#1}{\urlprefix }}%
\providecommand \urlprefix  [0]{URL }%
\providecommand \Eprint [0]{\href }%
\providecommand \doibase [0]{http://dx.doi.org/}%
\providecommand \selectlanguage [0]{\@gobble}%
\providecommand \bibinfo  [0]{\@secondoftwo}%
\providecommand \bibfield  [0]{\@secondoftwo}%
\providecommand \translation [1]{[#1]}%
\providecommand \BibitemOpen [0]{}%
\providecommand \bibitemStop [0]{}%
\providecommand \bibitemNoStop [0]{.\EOS\space}%
\providecommand \EOS [0]{\spacefactor3000\relax}%
\providecommand \BibitemShut  [1]{\csname bibitem#1\endcsname}%
\let\auto@bib@innerbib\@empty
\bibitem [{\citenamefont {Zyla}\ \emph {et~al.}(2020)\citenamefont {Zyla} \emph
  {et~al.}}]{pdg}%
  \BibitemOpen
  \bibfield  {author} {\bibinfo {author} {\bibfnamefont {P.~A.}\ \bibnamefont
  {Zyla}} \emph {et~al.} (\bibinfo {collaboration} {Particle Data Group}),\
  }\href {\doibase 10.1093/ptep/ptaa104} {\bibfield  {journal} {\bibinfo
  {journal} {PTEP}\ }\textbf {\bibinfo {volume} {2020}},\ \bibinfo {pages}
  {083C01} (\bibinfo {year} {2020})}\BibitemShut {NoStop}%
\bibitem [{\citenamefont {Esposito}\ \emph {et~al.}(2017)\citenamefont
  {Esposito}, \citenamefont {Pilloni},\ and\ \citenamefont
  {Polosa}}]{Esposito:2016noz}%
  \BibitemOpen
  \bibfield  {author} {\bibinfo {author} {\bibfnamefont {A.}~\bibnamefont
  {Esposito}}, \bibinfo {author} {\bibfnamefont {A.}~\bibnamefont {Pilloni}}, \
  and\ \bibinfo {author} {\bibfnamefont {A.~D.}\ \bibnamefont {Polosa}},\
  }\href {\doibase 10.1016/j.physrep.2016.11.002} {\bibfield  {journal}
  {\bibinfo  {journal} {Phys.Rept.}\ }\textbf {\bibinfo {volume} {668}},\
  \bibinfo {pages} {1} (\bibinfo {year} {2017})},\ \Eprint
  {http://arxiv.org/abs/1611.07920} {arXiv:1611.07920 [hep-ph]} \BibitemShut
  {NoStop}%
\bibitem [{\citenamefont {Olsen}\ \emph {et~al.}(2018)\citenamefont {Olsen},
  \citenamefont {Skwarnicki},\ and\ \citenamefont {Zieminska}}]{Olsen:2017bmm}%
  \BibitemOpen
  \bibfield  {author} {\bibinfo {author} {\bibfnamefont {S.~L.}\ \bibnamefont
  {Olsen}}, \bibinfo {author} {\bibfnamefont {T.}~\bibnamefont {Skwarnicki}}, \
  and\ \bibinfo {author} {\bibfnamefont {D.}~\bibnamefont {Zieminska}},\ }\href
  {\doibase 10.1103/RevModPhys.90.015003} {\bibfield  {journal} {\bibinfo
  {journal} {Rev.Mod.Phys.}\ }\textbf {\bibinfo {volume} {90}},\ \bibinfo
  {pages} {015003} (\bibinfo {year} {2018})},\ \Eprint
  {http://arxiv.org/abs/1708.04012} {arXiv:1708.04012 [hep-ph]} \BibitemShut
  {NoStop}%
\bibitem [{\citenamefont {Guo}\ \emph {et~al.}(2018)\citenamefont {Guo},
  \citenamefont {Hanhart}, \citenamefont {Mei{\ss}ner}, \citenamefont {Wang},
  \citenamefont {Zhao},\ and\ \citenamefont {Zou}}]{Guo:2017jvc}%
  \BibitemOpen
  \bibfield  {author} {\bibinfo {author} {\bibfnamefont {F.-K.}\ \bibnamefont
  {Guo}}, \bibinfo {author} {\bibfnamefont {C.}~\bibnamefont {Hanhart}},
  \bibinfo {author} {\bibfnamefont {U.-G.}\ \bibnamefont {Mei{\ss}ner}},
  \bibinfo {author} {\bibfnamefont {Q.}~\bibnamefont {Wang}}, \bibinfo {author}
  {\bibfnamefont {Q.}~\bibnamefont {Zhao}}, \ and\ \bibinfo {author}
  {\bibfnamefont {B.-S.}\ \bibnamefont {Zou}},\ }\href {\doibase
  10.1103/RevModPhys.90.015004} {\bibfield  {journal} {\bibinfo  {journal}
  {Rev.Mod.Phys.}\ }\textbf {\bibinfo {volume} {90}},\ \bibinfo {pages}
  {015004} (\bibinfo {year} {2018})},\ \Eprint
  {http://arxiv.org/abs/1705.00141} {arXiv:1705.00141 [hep-ph]} \BibitemShut
  {NoStop}%
\bibitem [{\citenamefont {Lebed}\ \emph {et~al.}(2017)\citenamefont {Lebed},
  \citenamefont {Mitchell},\ and\ \citenamefont {Swanson}}]{Lebed:2016hpi}%
  \BibitemOpen
  \bibfield  {author} {\bibinfo {author} {\bibfnamefont {R.~F.}\ \bibnamefont
  {Lebed}}, \bibinfo {author} {\bibfnamefont {R.~E.}\ \bibnamefont {Mitchell}},
  \ and\ \bibinfo {author} {\bibfnamefont {E.~S.}\ \bibnamefont {Swanson}},\
  }\href {\doibase 10.1016/j.ppnp.2016.11.003} {\bibfield  {journal} {\bibinfo
  {journal} {Prog.Part.Nucl.Phys.}\ }\textbf {\bibinfo {volume} {93}},\
  \bibinfo {pages} {143} (\bibinfo {year} {2017})},\ \Eprint
  {http://arxiv.org/abs/1610.04528} {arXiv:1610.04528 [hep-ph]} \BibitemShut
  {NoStop}%
\bibitem [{\citenamefont {Karliner}\ \emph {et~al.}(2018)\citenamefont
  {Karliner}, \citenamefont {Rosner},\ and\ \citenamefont
  {Skwarnicki}}]{Karliner:2017qhf}%
  \BibitemOpen
  \bibfield  {author} {\bibinfo {author} {\bibfnamefont {M.}~\bibnamefont
  {Karliner}}, \bibinfo {author} {\bibfnamefont {J.~L.}\ \bibnamefont
  {Rosner}}, \ and\ \bibinfo {author} {\bibfnamefont {T.}~\bibnamefont
  {Skwarnicki}},\ }\href {\doibase 10.1146/annurev-nucl-101917-020902}
  {\bibfield  {journal} {\bibinfo  {journal} {Ann.Rev.Nucl.Part.Sci}\ }\textbf
  {\bibinfo {volume} {68}},\ \bibinfo {pages} {17} (\bibinfo {year} {2018})},\
  \Eprint {http://arxiv.org/abs/1711.10626} {arXiv:1711.10626 [hep-ph]}
  \BibitemShut {NoStop}%
\bibitem [{\citenamefont {Guo}\ \emph {et~al.}(2020)\citenamefont {Guo},
  \citenamefont {Liu},\ and\ \citenamefont {Sakai}}]{Guo:2019twa}%
  \BibitemOpen
  \bibfield  {author} {\bibinfo {author} {\bibfnamefont {F.-K.}\ \bibnamefont
  {Guo}}, \bibinfo {author} {\bibfnamefont {X.-H.}\ \bibnamefont {Liu}}, \ and\
  \bibinfo {author} {\bibfnamefont {S.}~\bibnamefont {Sakai}},\ }\href
  {\doibase 10.1016/j.ppnp.2020.103757} {\bibfield  {journal} {\bibinfo
  {journal} {Prog.Part.Nucl.Phys.}\ }\textbf {\bibinfo {volume} {112}},\
  \bibinfo {pages} {103757} (\bibinfo {year} {2020})},\ \Eprint
  {http://arxiv.org/abs/1912.07030} {arXiv:1912.07030 [hep-ph]} \BibitemShut
  {NoStop}%
\bibitem [{\citenamefont {Ali}\ \emph {et~al.}(2019)\citenamefont {Ali},
  \citenamefont {Maiani},\ and\ \citenamefont {Polosa}}]{ali2019multiquark}%
  \BibitemOpen
  \bibfield  {author} {\bibinfo {author} {\bibfnamefont {A.}~\bibnamefont
  {Ali}}, \bibinfo {author} {\bibfnamefont {L.}~\bibnamefont {Maiani}}, \ and\
  \bibinfo {author} {\bibfnamefont {A.~D.}\ \bibnamefont {Polosa}},\ }\href
  {\doibase 10.1017/9781316761465} {\emph {\bibinfo {title} {Multiquark
  Hadrons}}}\ (\bibinfo  {publisher} {Cambridge University Press},\ \bibinfo
  {year} {2019})\BibitemShut {NoStop}%
\bibitem [{\citenamefont {Brambilla}\ \emph {et~al.}(2020)\citenamefont
  {Brambilla}, \citenamefont {Eidelman}, \citenamefont {Hanhart}, \citenamefont
  {Nefediev}, \citenamefont {Shen}, \citenamefont {Thomas}, \citenamefont
  {Vairo},\ and\ \citenamefont {Yuan}}]{Brambilla:2019esw}%
  \BibitemOpen
  \bibfield  {author} {\bibinfo {author} {\bibfnamefont {N.}~\bibnamefont
  {Brambilla}}, \bibinfo {author} {\bibfnamefont {S.}~\bibnamefont {Eidelman}},
  \bibinfo {author} {\bibfnamefont {C.}~\bibnamefont {Hanhart}}, \bibinfo
  {author} {\bibfnamefont {A.}~\bibnamefont {Nefediev}}, \bibinfo {author}
  {\bibfnamefont {C.-P.}\ \bibnamefont {Shen}}, \bibinfo {author}
  {\bibfnamefont {C.~E.}\ \bibnamefont {Thomas}}, \bibinfo {author}
  {\bibfnamefont {A.}~\bibnamefont {Vairo}}, \ and\ \bibinfo {author}
  {\bibfnamefont {C.-Z.}\ \bibnamefont {Yuan}},\ }\href {\doibase
  10.1016/j.physrep.2020.05.001} {\bibfield  {journal} {\bibinfo  {journal}
  {Phys.Rept.}\ }\textbf {\bibinfo {volume} {873}},\ \bibinfo {pages} {1}
  (\bibinfo {year} {2020})},\ \Eprint {http://arxiv.org/abs/1907.07583}
  {arXiv:1907.07583 [hep-ex]} \BibitemShut {NoStop}%
\bibitem [{\citenamefont {Caprini}\ \emph {et~al.}(2006)\citenamefont
  {Caprini}, \citenamefont {Colangelo},\ and\ \citenamefont
  {Leutwyler}}]{Caprini:2005zr}%
  \BibitemOpen
  \bibfield  {author} {\bibinfo {author} {\bibfnamefont {I.}~\bibnamefont
  {Caprini}}, \bibinfo {author} {\bibfnamefont {G.}~\bibnamefont {Colangelo}},
  \ and\ \bibinfo {author} {\bibfnamefont {H.}~\bibnamefont {Leutwyler}},\
  }\href {\doibase 10.1103/PhysRevLett.96.132001} {\bibfield  {journal}
  {\bibinfo  {journal} {Phys.Rev.Lett.}\ }\textbf {\bibinfo {volume} {96}},\
  \bibinfo {pages} {132001} (\bibinfo {year} {2006})},\ \Eprint
  {http://arxiv.org/abs/hep-ph/0512364} {arXiv:hep-ph/0512364 [hep-ph]}
  \BibitemShut {NoStop}%
\bibitem [{\citenamefont {Descotes-Genon}\ and\ \citenamefont
  {Moussallam}(2006)}]{DescotesGenon:2006uk}%
  \BibitemOpen
  \bibfield  {author} {\bibinfo {author} {\bibfnamefont {S.}~\bibnamefont
  {Descotes-Genon}}\ and\ \bibinfo {author} {\bibfnamefont {B.}~\bibnamefont
  {Moussallam}},\ }\href {\doibase 10.1140/epjc/s10052-006-0036-2} {\bibfield
  {journal} {\bibinfo  {journal} {Eur.Phys.J.}\ }\textbf {\bibinfo {volume}
  {C48}},\ \bibinfo {pages} {553} (\bibinfo {year} {2006})},\ \Eprint
  {http://arxiv.org/abs/hep-ph/0607133} {arXiv:hep-ph/0607133 [hep-ph]}
  \BibitemShut {NoStop}%
\bibitem [{\citenamefont {Garc\'ia-Mart\'in}\ \emph {et~al.}(2011)\citenamefont
  {Garc\'ia-Mart\'in}, \citenamefont {Kaminski}, \citenamefont {Pel\'aez},\
  and\ \citenamefont {Ruiz~de Elvira}}]{GarciaMartin:2011nna}%
  \BibitemOpen
  \bibfield  {author} {\bibinfo {author} {\bibfnamefont {R.}~\bibnamefont
  {Garc\'ia-Mart\'in}}, \bibinfo {author} {\bibfnamefont {R.}~\bibnamefont
  {Kaminski}}, \bibinfo {author} {\bibfnamefont {J.~R.}\ \bibnamefont
  {Pel\'aez}}, \ and\ \bibinfo {author} {\bibfnamefont {J.}~\bibnamefont
  {Ruiz~de Elvira}},\ }\href {\doibase 10.1103/PhysRevLett.107.072001}
  {\bibfield  {journal} {\bibinfo  {journal} {Phys.Rev.Lett.}\ }\textbf
  {\bibinfo {volume} {107}},\ \bibinfo {pages} {072001} (\bibinfo {year}
  {2011})},\ \Eprint {http://arxiv.org/abs/1107.1635} {arXiv:1107.1635
  [hep-ph]} \BibitemShut {NoStop}%
\bibitem [{\citenamefont {Hoferichter}\ \emph {et~al.}(2011)\citenamefont
  {Hoferichter}, \citenamefont {Phillips},\ and\ \citenamefont
  {Schat}}]{Hoferichter:2011wk}%
  \BibitemOpen
  \bibfield  {author} {\bibinfo {author} {\bibfnamefont {M.}~\bibnamefont
  {Hoferichter}}, \bibinfo {author} {\bibfnamefont {D.~R.}\ \bibnamefont
  {Phillips}}, \ and\ \bibinfo {author} {\bibfnamefont {C.}~\bibnamefont
  {Schat}},\ }\href {\doibase 10.1140/epjc/s10052-011-1743-x} {\bibfield
  {journal} {\bibinfo  {journal} {Eur.Phys.J.}\ }\textbf {\bibinfo {volume}
  {C71}},\ \bibinfo {pages} {1743} (\bibinfo {year} {2011})},\ \Eprint
  {http://arxiv.org/abs/1106.4147} {arXiv:1106.4147 [hep-ph]} \BibitemShut
  {NoStop}%
\bibitem [{\citenamefont {Moussallam}(2011)}]{Moussallam:2011zg}%
  \BibitemOpen
  \bibfield  {author} {\bibinfo {author} {\bibfnamefont {B.}~\bibnamefont
  {Moussallam}},\ }\href {\doibase 10.1140/epjc/s10052-011-1814-z} {\bibfield
  {journal} {\bibinfo  {journal} {Eur.Phys.J.}\ }\textbf {\bibinfo {volume}
  {C71}},\ \bibinfo {pages} {1814} (\bibinfo {year} {2011})},\ \Eprint
  {http://arxiv.org/abs/1110.6074} {arXiv:1110.6074 [hep-ph]} \BibitemShut
  {NoStop}%
\bibitem [{\citenamefont {Ditsche}\ \emph {et~al.}(2012)\citenamefont
  {Ditsche}, \citenamefont {Hoferichter}, \citenamefont {Kubis},\ and\
  \citenamefont {Mei{\ss}ner}}]{Ditsche:2012fv}%
  \BibitemOpen
  \bibfield  {author} {\bibinfo {author} {\bibfnamefont {C.}~\bibnamefont
  {Ditsche}}, \bibinfo {author} {\bibfnamefont {M.}~\bibnamefont
  {Hoferichter}}, \bibinfo {author} {\bibfnamefont {B.}~\bibnamefont {Kubis}},
  \ and\ \bibinfo {author} {\bibfnamefont {U.-G.}\ \bibnamefont
  {Mei{\ss}ner}},\ }\href {\doibase 10.1007/JHEP06(2012)043} {\bibfield
  {journal} {\bibinfo  {journal} {JHEP}\ }\textbf {\bibinfo {volume} {06}},\
  \bibinfo {pages} {043} (\bibinfo {year} {2012})},\ \Eprint
  {http://arxiv.org/abs/1203.4758} {arXiv:1203.4758 [hep-ph]} \BibitemShut
  {NoStop}%
\bibitem [{\citenamefont {Pel\'aez}\ and\ \citenamefont
  {Rodas}(2020{\natexlab{a}})}]{Pelaez:2020uiw}%
  \BibitemOpen
  \bibfield  {author} {\bibinfo {author} {\bibfnamefont {J.}~\bibnamefont
  {Pel\'aez}}\ and\ \bibinfo {author} {\bibfnamefont {A.}~\bibnamefont
  {Rodas}},\ }\href {\doibase 10.1103/PhysRevLett.124.172001} {\bibfield
  {journal} {\bibinfo  {journal} {Phys.Rev.Lett.}\ }\textbf {\bibinfo {volume}
  {124}},\ \bibinfo {pages} {172001} (\bibinfo {year} {2020}{\natexlab{a}})},\
  \Eprint {http://arxiv.org/abs/2001.08153} {arXiv:2001.08153 [hep-ph]}
  \BibitemShut {NoStop}%
\bibitem [{\citenamefont {Pel\'aez}\ and\ \citenamefont
  {Rodas}(2020{\natexlab{b}})}]{Pelaez:2020gnd}%
  \BibitemOpen
  \bibfield  {author} {\bibinfo {author} {\bibfnamefont {J.}~\bibnamefont
  {Pel\'aez}}\ and\ \bibinfo {author} {\bibfnamefont {A.}~\bibnamefont
  {Rodas}},\ }\href@noop {} {\enquote {\bibinfo {title} {{Dispersive $\pi K\to
  \pi K$ and $\pi \pi \to K \bar{K}$ amplitudes from scattering data, threshold
  parameters and the lightest strange resonance $\kappa$ or $K^*_0(700)$}},}\ }
  (\bibinfo {year} {2020}{\natexlab{b}}),\ \Eprint
  {http://arxiv.org/abs/2010.11222} {arXiv:2010.11222 [hep-ph]} \BibitemShut
  {NoStop}%
\bibitem [{\citenamefont {Albaladejo}\ \emph {et~al.}(2021)\citenamefont
  {Albaladejo} \emph {et~al.}}]{JPAC:2021rxu}%
  \BibitemOpen
  \bibfield  {author} {\bibinfo {author} {\bibfnamefont {M.}~\bibnamefont
  {Albaladejo}} \emph {et~al.} (\bibinfo {collaboration} {JPAC}),\ }\href@noop
  {} {\enquote {\bibinfo {title} {{Novel approaches in Hadron Spectroscopy}},}\
  } (\bibinfo {year} {2021}),\ \Eprint {http://arxiv.org/abs/2112.13436}
  {arXiv:2112.13436 [hep-ph]} \BibitemShut {NoStop}%
\bibitem [{\citenamefont {Masjuan}\ and\ \citenamefont
  {Sanz-Cillero}(2013)}]{Masjuan:2013jha}%
  \BibitemOpen
  \bibfield  {author} {\bibinfo {author} {\bibfnamefont {P.}~\bibnamefont
  {Masjuan}}\ and\ \bibinfo {author} {\bibfnamefont {J.~J.}\ \bibnamefont
  {Sanz-Cillero}},\ }\href {\doibase 10.1140/epjc/s10052-013-2594-4} {\bibfield
   {journal} {\bibinfo  {journal} {Eur.Phys.J.}\ }\textbf {\bibinfo {volume}
  {C73}},\ \bibinfo {pages} {2594} (\bibinfo {year} {2013})},\ \Eprint
  {http://arxiv.org/abs/1306.6308} {arXiv:1306.6308 [hep-ph]} \BibitemShut
  {NoStop}%
\bibitem [{\citenamefont {Masjuan}\ \emph {et~al.}(2014)\citenamefont
  {Masjuan}, \citenamefont {Ruiz~de Elvira},\ and\ \citenamefont
  {Sanz-Cillero}}]{Masjuan:2014psa}%
  \BibitemOpen
  \bibfield  {author} {\bibinfo {author} {\bibfnamefont {P.}~\bibnamefont
  {Masjuan}}, \bibinfo {author} {\bibfnamefont {J.}~\bibnamefont {Ruiz~de
  Elvira}}, \ and\ \bibinfo {author} {\bibfnamefont {J.~J.}\ \bibnamefont
  {Sanz-Cillero}},\ }\href {\doibase 10.1103/PhysRevD.90.097901} {\bibfield
  {journal} {\bibinfo  {journal} {Phys.Rev.}\ }\textbf {\bibinfo {volume}
  {D90}},\ \bibinfo {pages} {097901} (\bibinfo {year} {2014})},\ \Eprint
  {http://arxiv.org/abs/1410.2397} {arXiv:1410.2397 [hep-ph]} \BibitemShut
  {NoStop}%
\bibitem [{\citenamefont {Caprini}\ \emph {et~al.}(2016)\citenamefont
  {Caprini}, \citenamefont {Masjuan}, \citenamefont {Ruiz~de Elvira},\ and\
  \citenamefont {Sanz-Cillero}}]{Caprini:2016uxy}%
  \BibitemOpen
  \bibfield  {author} {\bibinfo {author} {\bibfnamefont {I.}~\bibnamefont
  {Caprini}}, \bibinfo {author} {\bibfnamefont {P.}~\bibnamefont {Masjuan}},
  \bibinfo {author} {\bibfnamefont {J.}~\bibnamefont {Ruiz~de Elvira}}, \ and\
  \bibinfo {author} {\bibfnamefont {J.~J.}\ \bibnamefont {Sanz-Cillero}},\
  }\href {\doibase 10.1103/PhysRevD.93.076004} {\bibfield  {journal} {\bibinfo
  {journal} {Phys.Rev.}\ }\textbf {\bibinfo {volume} {D93}},\ \bibinfo {pages}
  {076004} (\bibinfo {year} {2016})},\ \Eprint
  {http://arxiv.org/abs/1602.02062} {arXiv:1602.02062 [hep-ph]} \BibitemShut
  {NoStop}%
\bibitem [{\citenamefont {Pel\'aez}\ \emph {et~al.}(2017)\citenamefont
  {Pel\'aez}, \citenamefont {Rodas},\ and\ \citenamefont {Ruiz~de
  Elvira}}]{Pelaez:2016klv}%
  \BibitemOpen
  \bibfield  {author} {\bibinfo {author} {\bibfnamefont {J.~R.}\ \bibnamefont
  {Pel\'aez}}, \bibinfo {author} {\bibfnamefont {A.}~\bibnamefont {Rodas}}, \
  and\ \bibinfo {author} {\bibfnamefont {J.}~\bibnamefont {Ruiz~de Elvira}},\
  }\href {\doibase 10.1140/epjc/s10052-017-4668-1} {\bibfield  {journal}
  {\bibinfo  {journal} {Eur.Phys.J.}\ }\textbf {\bibinfo {volume} {C77}},\
  \bibinfo {pages} {91} (\bibinfo {year} {2017})},\ \Eprint
  {http://arxiv.org/abs/1612.07966} {arXiv:1612.07966 [hep-ph]} \BibitemShut
  {NoStop}%
\bibitem [{\citenamefont {\v{S}varc}\ \emph {et~al.}(2013)\citenamefont
  {\v{S}varc}, \citenamefont {Hadzimehmedovic}, \citenamefont {Osmanovic},
  \citenamefont {Stahov}, \citenamefont {Tiator},\ and\ \citenamefont
  {Workman}}]{Svarc:2013laa}%
  \BibitemOpen
  \bibfield  {author} {\bibinfo {author} {\bibfnamefont {A.}~\bibnamefont
  {\v{S}varc}}, \bibinfo {author} {\bibfnamefont {M.}~\bibnamefont
  {Hadzimehmedovic}}, \bibinfo {author} {\bibfnamefont {H.}~\bibnamefont
  {Osmanovic}}, \bibinfo {author} {\bibfnamefont {J.}~\bibnamefont {Stahov}},
  \bibinfo {author} {\bibfnamefont {L.}~\bibnamefont {Tiator}}, \ and\ \bibinfo
  {author} {\bibfnamefont {R.~L.}\ \bibnamefont {Workman}},\ }\href {\doibase
  10.1103/PhysRevC.88.035206} {\bibfield  {journal} {\bibinfo  {journal}
  {Phys.Rev.}\ }\textbf {\bibinfo {volume} {C88}},\ \bibinfo {pages} {035206}
  (\bibinfo {year} {2013})},\ \Eprint {http://arxiv.org/abs/1307.4613}
  {arXiv:1307.4613 [hep-ph]} \BibitemShut {NoStop}%
\bibitem [{\citenamefont {\v{S}varc}\ \emph {et~al.}(2014)\citenamefont
  {\v{S}varc}, \citenamefont {Had\v{z}imehmedovi\'c}, \citenamefont
  {Osmanovi\'c}, \citenamefont {Stahov}, \citenamefont {Tiator},\ and\
  \citenamefont {Workman}}]{Svarc:2014sqa}%
  \BibitemOpen
  \bibfield  {author} {\bibinfo {author} {\bibfnamefont {A.}~\bibnamefont
  {\v{S}varc}}, \bibinfo {author} {\bibfnamefont {M.}~\bibnamefont
  {Had\v{z}imehmedovi\'c}}, \bibinfo {author} {\bibfnamefont {H.}~\bibnamefont
  {Osmanovi\'c}}, \bibinfo {author} {\bibfnamefont {J.}~\bibnamefont {Stahov}},
  \bibinfo {author} {\bibfnamefont {L.}~\bibnamefont {Tiator}}, \ and\ \bibinfo
  {author} {\bibfnamefont {R.~L.}\ \bibnamefont {Workman}},\ }\href {\doibase
  10.1103/PhysRevC.89.065208} {\bibfield  {journal} {\bibinfo  {journal}
  {Phys.Rev.}\ }\textbf {\bibinfo {volume} {C89}},\ \bibinfo {pages} {065208}
  (\bibinfo {year} {2014})},\ \Eprint {http://arxiv.org/abs/1404.1544}
  {arXiv:1404.1544 [nucl-th]} \BibitemShut {NoStop}%
\bibitem [{\citenamefont {Schlessinger}(1968)}]{PhysRev.167.1411}%
  \BibitemOpen
  \bibfield  {author} {\bibinfo {author} {\bibfnamefont {L.}~\bibnamefont
  {Schlessinger}},\ }\href {\doibase 10.1103/PhysRev.167.1411} {\bibfield
  {journal} {\bibinfo  {journal} {Phys.Rev.}\ }\textbf {\bibinfo {volume}
  {167}},\ \bibinfo {pages} {1411} (\bibinfo {year} {1968})}\BibitemShut
  {NoStop}%
\bibitem [{\citenamefont {Schlessinger}\ and\ \citenamefont
  {Schwartz}(1966)}]{Schlessinger:1966zz}%
  \BibitemOpen
  \bibfield  {author} {\bibinfo {author} {\bibfnamefont {L.}~\bibnamefont
  {Schlessinger}}\ and\ \bibinfo {author} {\bibfnamefont {C.}~\bibnamefont
  {Schwartz}},\ }\href {\doibase 10.1103/PhysRevLett.16.1173} {\bibfield
  {journal} {\bibinfo  {journal} {Phys.Rev.Lett.}\ }\textbf {\bibinfo {volume}
  {16}},\ \bibinfo {pages} {1173} (\bibinfo {year} {1966})}\BibitemShut
  {NoStop}%
\bibitem [{\citenamefont {Tripolt}\ \emph {et~al.}(2017)\citenamefont
  {Tripolt}, \citenamefont {Haritan}, \citenamefont {Wambach},\ and\
  \citenamefont {Moiseyev}}]{Tripolt:2016cya}%
  \BibitemOpen
  \bibfield  {author} {\bibinfo {author} {\bibfnamefont {R.-A.}\ \bibnamefont
  {Tripolt}}, \bibinfo {author} {\bibfnamefont {I.}~\bibnamefont {Haritan}},
  \bibinfo {author} {\bibfnamefont {J.}~\bibnamefont {Wambach}}, \ and\
  \bibinfo {author} {\bibfnamefont {N.}~\bibnamefont {Moiseyev}},\ }\href
  {\doibase 10.1016/j.physletb.2017.10.001} {\bibfield  {journal} {\bibinfo
  {journal} {Phys.Lett.}\ }\textbf {\bibinfo {volume} {B774}},\ \bibinfo
  {pages} {411} (\bibinfo {year} {2017})},\ \Eprint
  {http://arxiv.org/abs/1610.03252} {arXiv:1610.03252 [hep-ph]} \BibitemShut
  {NoStop}%
\bibitem [{\citenamefont {Tripolt}\ \emph {et~al.}(2019)\citenamefont
  {Tripolt}, \citenamefont {Gubler}, \citenamefont {Ulybyshev},\ and\
  \citenamefont {Von~Smekal}}]{Tripolt:2018xeo}%
  \BibitemOpen
  \bibfield  {author} {\bibinfo {author} {\bibfnamefont {R.-A.}\ \bibnamefont
  {Tripolt}}, \bibinfo {author} {\bibfnamefont {P.}~\bibnamefont {Gubler}},
  \bibinfo {author} {\bibfnamefont {M.}~\bibnamefont {Ulybyshev}}, \ and\
  \bibinfo {author} {\bibfnamefont {L.}~\bibnamefont {Von~Smekal}},\ }\href
  {\doibase 10.1016/j.cpc.2018.11.012} {\bibfield  {journal} {\bibinfo
  {journal} {Comput.Phys.Commun.}\ }\textbf {\bibinfo {volume} {237}},\
  \bibinfo {pages} {129} (\bibinfo {year} {2019})},\ \Eprint
  {http://arxiv.org/abs/1801.10348} {arXiv:1801.10348 [hep-ph]} \BibitemShut
  {NoStop}%
\bibitem [{\citenamefont {Chen}\ \emph {et~al.}(2019)\citenamefont {Chen},
  \citenamefont {Lu}, \citenamefont {Binosi}, \citenamefont {Roberts},
  \citenamefont {Rodr\'\i{}guez-Quintero},\ and\ \citenamefont
  {Segovia}}]{Chen:2018nsg}%
  \BibitemOpen
  \bibfield  {author} {\bibinfo {author} {\bibfnamefont {C.}~\bibnamefont
  {Chen}}, \bibinfo {author} {\bibfnamefont {Y.}~\bibnamefont {Lu}}, \bibinfo
  {author} {\bibfnamefont {D.}~\bibnamefont {Binosi}}, \bibinfo {author}
  {\bibfnamefont {C.~D.}\ \bibnamefont {Roberts}}, \bibinfo {author}
  {\bibfnamefont {J.}~\bibnamefont {Rodr\'\i{}guez-Quintero}}, \ and\ \bibinfo
  {author} {\bibfnamefont {J.}~\bibnamefont {Segovia}},\ }\href {\doibase
  10.1103/PhysRevD.99.034013} {\bibfield  {journal} {\bibinfo  {journal}
  {Phys.Rev.}\ }\textbf {\bibinfo {volume} {D99}},\ \bibinfo {pages} {034013}
  (\bibinfo {year} {2019})},\ \Eprint {http://arxiv.org/abs/1811.08440}
  {arXiv:1811.08440 [nucl-th]} \BibitemShut {NoStop}%
\bibitem [{\citenamefont {Binosi}\ \emph {et~al.}(2019)\citenamefont {Binosi},
  \citenamefont {Chang}, \citenamefont {Ding}, \citenamefont {Gao},
  \citenamefont {Papavassiliou},\ and\ \citenamefont
  {Roberts}}]{Binosi:2018rht}%
  \BibitemOpen
  \bibfield  {author} {\bibinfo {author} {\bibfnamefont {D.}~\bibnamefont
  {Binosi}}, \bibinfo {author} {\bibfnamefont {L.}~\bibnamefont {Chang}},
  \bibinfo {author} {\bibfnamefont {M.}~\bibnamefont {Ding}}, \bibinfo {author}
  {\bibfnamefont {F.}~\bibnamefont {Gao}}, \bibinfo {author} {\bibfnamefont
  {J.}~\bibnamefont {Papavassiliou}}, \ and\ \bibinfo {author} {\bibfnamefont
  {C.~D.}\ \bibnamefont {Roberts}},\ }\href {\doibase
  10.1016/j.physletb.2019.01.033} {\bibfield  {journal} {\bibinfo  {journal}
  {Phys.Lett.}\ }\textbf {\bibinfo {volume} {B790}},\ \bibinfo {pages} {257}
  (\bibinfo {year} {2019})},\ \Eprint {http://arxiv.org/abs/1812.05112}
  {arXiv:1812.05112 [nucl-th]} \BibitemShut {NoStop}%
\bibitem [{\citenamefont {Binosi}\ and\ \citenamefont
  {Tripolt}(2020)}]{Binosi:2019ecz}%
  \BibitemOpen
  \bibfield  {author} {\bibinfo {author} {\bibfnamefont {D.}~\bibnamefont
  {Binosi}}\ and\ \bibinfo {author} {\bibfnamefont {R.-A.}\ \bibnamefont
  {Tripolt}},\ }\href {\doibase 10.1016/j.physletb.2019.135171} {\bibfield
  {journal} {\bibinfo  {journal} {Phys.Lett.}\ }\textbf {\bibinfo {volume}
  {B801}},\ \bibinfo {pages} {135171} (\bibinfo {year} {2020})},\ \Eprint
  {http://arxiv.org/abs/1904.08172} {arXiv:1904.08172 [hep-ph]} \BibitemShut
  {NoStop}%
\bibitem [{\citenamefont {Eichmann}\ \emph {et~al.}(2019)\citenamefont
  {Eichmann}, \citenamefont {Duarte}, \citenamefont {Pe\~na},\ and\
  \citenamefont {Stadler}}]{Eichmann:2019dts}%
  \BibitemOpen
  \bibfield  {author} {\bibinfo {author} {\bibfnamefont {G.}~\bibnamefont
  {Eichmann}}, \bibinfo {author} {\bibfnamefont {P.}~\bibnamefont {Duarte}},
  \bibinfo {author} {\bibfnamefont {M.~T.}\ \bibnamefont {Pe\~na}}, \ and\
  \bibinfo {author} {\bibfnamefont {A.}~\bibnamefont {Stadler}},\ }\href
  {\doibase 10.1103/PhysRevD.100.094001} {\bibfield  {journal} {\bibinfo
  {journal} {Phys.Rev.}\ }\textbf {\bibinfo {volume} {D100}},\ \bibinfo {pages}
  {094001} (\bibinfo {year} {2019})},\ \Eprint
  {http://arxiv.org/abs/1907.05402} {arXiv:1907.05402 [hep-ph]} \BibitemShut
  {NoStop}%
\bibitem [{\citenamefont {Yao}\ \emph {et~al.}(2020)\citenamefont {Yao},
  \citenamefont {Binosi}, \citenamefont {Cui}, \citenamefont {Roberts},
  \citenamefont {Xu},\ and\ \citenamefont {Zong}}]{Yao:2020vef}%
  \BibitemOpen
  \bibfield  {author} {\bibinfo {author} {\bibfnamefont {Z.-Q.}\ \bibnamefont
  {Yao}}, \bibinfo {author} {\bibfnamefont {D.}~\bibnamefont {Binosi}},
  \bibinfo {author} {\bibfnamefont {Z.-F.}\ \bibnamefont {Cui}}, \bibinfo
  {author} {\bibfnamefont {C.~D.}\ \bibnamefont {Roberts}}, \bibinfo {author}
  {\bibfnamefont {S.-S.}\ \bibnamefont {Xu}}, \ and\ \bibinfo {author}
  {\bibfnamefont {H.~S.}\ \bibnamefont {Zong}},\ }\href {\doibase
  10.1103/PhysRevD.102.014007} {\bibfield  {journal} {\bibinfo  {journal}
  {Phys.Rev.}\ }\textbf {\bibinfo {volume} {D102}},\ \bibinfo {pages} {014007}
  (\bibinfo {year} {2020})},\ \Eprint {http://arxiv.org/abs/2003.04420}
  {arXiv:2003.04420 [hep-ph]} \BibitemShut {NoStop}%
\bibitem [{\citenamefont {Yao}\ \emph {et~al.}(2021)\citenamefont {Yao},
  \citenamefont {Binosi}, \citenamefont {Cui},\ and\ \citenamefont
  {Roberts}}]{Yao:2021pyf}%
  \BibitemOpen
  \bibfield  {author} {\bibinfo {author} {\bibfnamefont {Z.-Q.}\ \bibnamefont
  {Yao}}, \bibinfo {author} {\bibfnamefont {D.}~\bibnamefont {Binosi}},
  \bibinfo {author} {\bibfnamefont {Z.-F.}\ \bibnamefont {Cui}}, \ and\
  \bibinfo {author} {\bibfnamefont {C.~D.}\ \bibnamefont {Roberts}},\ }\href
  {\doibase 10.1016/j.physletb.2021.136344} {\bibfield  {journal} {\bibinfo
  {journal} {Phys.Lett.}\ }\textbf {\bibinfo {volume} {B818}},\ \bibinfo
  {pages} {136344} (\bibinfo {year} {2021})},\ \Eprint
  {http://arxiv.org/abs/2104.10261} {arXiv:2104.10261 [hep-ph]} \BibitemShut
  {NoStop}%
\bibitem [{\citenamefont {Cui}\ \emph {et~al.}(2021{\natexlab{a}})\citenamefont
  {Cui}, \citenamefont {Binosi}, \citenamefont {Roberts},\ and\ \citenamefont
  {Schmidt}}]{Cui:2021vgm}%
  \BibitemOpen
  \bibfield  {author} {\bibinfo {author} {\bibfnamefont {Z.-F.}\ \bibnamefont
  {Cui}}, \bibinfo {author} {\bibfnamefont {D.}~\bibnamefont {Binosi}},
  \bibinfo {author} {\bibfnamefont {C.~D.}\ \bibnamefont {Roberts}}, \ and\
  \bibinfo {author} {\bibfnamefont {S.~M.}\ \bibnamefont {Schmidt}},\ }\href
  {\doibase 10.1103/PhysRevLett.127.092001} {\bibfield  {journal} {\bibinfo
  {journal} {Phys.Rev.Lett.}\ }\textbf {\bibinfo {volume} {127}},\ \bibinfo
  {pages} {092001} (\bibinfo {year} {2021}{\natexlab{a}})},\ \Eprint
  {http://arxiv.org/abs/2102.01180} {arXiv:2102.01180 [hep-ph]} \BibitemShut
  {NoStop}%
\bibitem [{\citenamefont {Cui}\ \emph {et~al.}(2021{\natexlab{b}})\citenamefont
  {Cui}, \citenamefont {Binosi}, \citenamefont {Roberts},\ and\ \citenamefont
  {Schmidt}}]{Cui:2021aee}%
  \BibitemOpen
  \bibfield  {author} {\bibinfo {author} {\bibfnamefont {Z.-F.}\ \bibnamefont
  {Cui}}, \bibinfo {author} {\bibfnamefont {D.}~\bibnamefont {Binosi}},
  \bibinfo {author} {\bibfnamefont {C.~D.}\ \bibnamefont {Roberts}}, \ and\
  \bibinfo {author} {\bibfnamefont {S.~M.}\ \bibnamefont {Schmidt}},\ }\href
  {\doibase 10.1016/j.physletb.2021.136631} {\bibfield  {journal} {\bibinfo
  {journal} {Phys.Lett.}\ }\textbf {\bibinfo {volume} {B822}},\ \bibinfo
  {pages} {136631} (\bibinfo {year} {2021}{\natexlab{b}})},\ \Eprint
  {http://arxiv.org/abs/2108.04948} {arXiv:2108.04948 [hep-ph]} \BibitemShut
  {NoStop}%
\bibitem [{\citenamefont {Cui}\ \emph {et~al.}(2021{\natexlab{c}})\citenamefont
  {Cui}, \citenamefont {Binosi}, \citenamefont {Roberts},\ and\ \citenamefont
  {Schmidt}}]{Cui:2021skn}%
  \BibitemOpen
  \bibfield  {author} {\bibinfo {author} {\bibfnamefont {Z.-F.}\ \bibnamefont
  {Cui}}, \bibinfo {author} {\bibfnamefont {D.}~\bibnamefont {Binosi}},
  \bibinfo {author} {\bibfnamefont {C.~D.}\ \bibnamefont {Roberts}}, \ and\
  \bibinfo {author} {\bibfnamefont {S.~M.}\ \bibnamefont {Schmidt}},\ }\href
  {\doibase 10.1088/0256-307X/38/12/121401} {\bibfield  {journal} {\bibinfo
  {journal} {Chin.Phys.Lett.}\ }\textbf {\bibinfo {volume} {38}},\ \bibinfo
  {pages} {121401} (\bibinfo {year} {2021}{\natexlab{c}})},\ \Eprint
  {http://arxiv.org/abs/2109.08768} {arXiv:2109.08768 [hep-ph]} \BibitemShut
  {NoStop}%
\bibitem [{\citenamefont {Ablikim}\ \emph {et~al.}(2015)\citenamefont {Ablikim}
  \emph {et~al.}}]{BESIII:2015rug}%
  \BibitemOpen
  \bibfield  {author} {\bibinfo {author} {\bibfnamefont {M.}~\bibnamefont
  {Ablikim}} \emph {et~al.} (\bibinfo {collaboration} {BESIII}),\ }\href
  {\doibase 10.1103/PhysRevD.92.052003} {\bibfield  {journal} {\bibinfo
  {journal} {Phys.Rev.}\ }\textbf {\bibinfo {volume} {D92}},\ \bibinfo {pages}
  {052003} (\bibinfo {year} {2015})},\ \bibinfo {note} {[Erratum:
  Phys.Rev.D93,no.3,039906(2016)]},\ \Eprint {http://arxiv.org/abs/1506.00546}
  {arXiv:1506.00546 [hep-ex]} \BibitemShut {NoStop}%
\bibitem [{\citenamefont {Ochs}(2013)}]{Ochs:2013gi}%
  \BibitemOpen
  \bibfield  {author} {\bibinfo {author} {\bibfnamefont {W.}~\bibnamefont
  {Ochs}},\ }\href {\doibase 10.1088/0954-3899/40/4/043001} {\bibfield
  {journal} {\bibinfo  {journal} {J.Phys.}\ }\textbf {\bibinfo {volume}
  {G40}},\ \bibinfo {pages} {043001} (\bibinfo {year} {2013})},\ \Eprint
  {http://arxiv.org/abs/1301.5183} {arXiv:1301.5183 [hep-ph]} \BibitemShut
  {NoStop}%
\bibitem [{\citenamefont {Huber}\ \emph {et~al.}(2020)\citenamefont {Huber},
  \citenamefont {Fischer},\ and\ \citenamefont
  {Sanchis-Alepuz}}]{Huber:2020ngt}%
  \BibitemOpen
  \bibfield  {author} {\bibinfo {author} {\bibfnamefont {M.~Q.}\ \bibnamefont
  {Huber}}, \bibinfo {author} {\bibfnamefont {C.~S.}\ \bibnamefont {Fischer}},
  \ and\ \bibinfo {author} {\bibfnamefont {H.}~\bibnamefont {Sanchis-Alepuz}},\
  }\href {\doibase 10.1140/epjc/s10052-020-08649-6} {\bibfield  {journal}
  {\bibinfo  {journal} {Eur.Phys.J.}\ }\textbf {\bibinfo {volume} {C80}},\
  \bibinfo {pages} {1077} (\bibinfo {year} {2020})},\ \Eprint
  {http://arxiv.org/abs/2004.00415} {arXiv:2004.00415 [hep-ph]} \BibitemShut
  {NoStop}%
\bibitem [{\citenamefont {Rodas}\ \emph {et~al.}(2022)\citenamefont {Rodas},
  \citenamefont {Pilloni}, \citenamefont {Albaladejo}, \citenamefont
  {Fernandez-Ramirez}, \citenamefont {Mathieu},\ and\ \citenamefont
  {Szczepaniak}}]{Rodas:2021tyb}%
  \BibitemOpen
  \bibfield  {author} {\bibinfo {author} {\bibfnamefont {A.}~\bibnamefont
  {Rodas}}, \bibinfo {author} {\bibfnamefont {A.}~\bibnamefont {Pilloni}},
  \bibinfo {author} {\bibfnamefont {M.}~\bibnamefont {Albaladejo}}, \bibinfo
  {author} {\bibfnamefont {C.}~\bibnamefont {Fernandez-Ramirez}}, \bibinfo
  {author} {\bibfnamefont {V.}~\bibnamefont {Mathieu}}, \ and\ \bibinfo
  {author} {\bibfnamefont {A.~P.}\ \bibnamefont {Szczepaniak}} (\bibinfo
  {collaboration} {Joint Physics Analysis Center}),\ }\href {\doibase
  10.1140/epjc/s10052-022-10014-8} {\bibfield  {journal} {\bibinfo  {journal}
  {Eur.Phys.J.}\ }\textbf {\bibinfo {volume} {C82}},\ \bibinfo {pages} {80}
  (\bibinfo {year} {2022})},\ \Eprint {http://arxiv.org/abs/2110.00027}
  {arXiv:2110.00027 [hep-ph]} \BibitemShut {NoStop}%
\bibitem [{\citenamefont {Ropertz}\ \emph {et~al.}(2018)\citenamefont
  {Ropertz}, \citenamefont {Hanhart},\ and\ \citenamefont
  {Kubis}}]{Ropertz:2018stk}%
  \BibitemOpen
  \bibfield  {author} {\bibinfo {author} {\bibfnamefont {S.}~\bibnamefont
  {Ropertz}}, \bibinfo {author} {\bibfnamefont {C.}~\bibnamefont {Hanhart}}, \
  and\ \bibinfo {author} {\bibfnamefont {B.}~\bibnamefont {Kubis}},\ }\href
  {\doibase 10.1140/epjc/s10052-018-6416-6} {\bibfield  {journal} {\bibinfo
  {journal} {Eur.Phys.J.}\ }\textbf {\bibinfo {volume} {C78}},\ \bibinfo
  {pages} {1000} (\bibinfo {year} {2018})},\ \Eprint
  {http://arxiv.org/abs/1809.06867} {arXiv:1809.06867 [hep-ph]} \BibitemShut
  {NoStop}%
\bibitem [{\citenamefont {Sarantsev}\ \emph {et~al.}(2021)\citenamefont
  {Sarantsev}, \citenamefont {Denisenko}, \citenamefont {Thoma},\ and\
  \citenamefont {Klempt}}]{Sarantsev:2021ein}%
  \BibitemOpen
  \bibfield  {author} {\bibinfo {author} {\bibfnamefont {A.~V.}\ \bibnamefont
  {Sarantsev}}, \bibinfo {author} {\bibfnamefont {I.}~\bibnamefont
  {Denisenko}}, \bibinfo {author} {\bibfnamefont {U.}~\bibnamefont {Thoma}}, \
  and\ \bibinfo {author} {\bibfnamefont {E.}~\bibnamefont {Klempt}},\ }\href
  {\doibase 10.1016/j.physletb.2021.136227} {\bibfield  {journal} {\bibinfo
  {journal} {Phys.Lett.}\ }\textbf {\bibinfo {volume} {B816}},\ \bibinfo
  {pages} {136227} (\bibinfo {year} {2021})},\ \Eprint
  {http://arxiv.org/abs/2103.09680} {arXiv:2103.09680 [hep-ph]} \BibitemShut
  {NoStop}%
\bibitem [{\citenamefont {Ablikim}\ \emph {et~al.}(2018)\citenamefont {Ablikim}
  \emph {et~al.}}]{BESIII:2018ubj}%
  \BibitemOpen
  \bibfield  {author} {\bibinfo {author} {\bibfnamefont {M.}~\bibnamefont
  {Ablikim}} \emph {et~al.} (\bibinfo {collaboration} {BESIII}),\ }\href
  {\doibase 10.1103/PhysRevD.98.072003} {\bibfield  {journal} {\bibinfo
  {journal} {Phys.Rev.}\ }\textbf {\bibinfo {volume} {D98}},\ \bibinfo {pages}
  {072003} (\bibinfo {year} {2018})},\ \Eprint
  {http://arxiv.org/abs/1808.06946} {arXiv:1808.06946 [hep-ex]} \BibitemShut
  {NoStop}%
\bibitem [{\citenamefont {Castillejo}\ \emph {et~al.}(1956)\citenamefont
  {Castillejo}, \citenamefont {Dalitz},\ and\ \citenamefont
  {Dyson}}]{Castillejo:1955ed}%
  \BibitemOpen
  \bibfield  {author} {\bibinfo {author} {\bibfnamefont {L.}~\bibnamefont
  {Castillejo}}, \bibinfo {author} {\bibfnamefont {R.~H.}\ \bibnamefont
  {Dalitz}}, \ and\ \bibinfo {author} {\bibfnamefont {F.~J.}\ \bibnamefont
  {Dyson}},\ }\href {\doibase 10.1103/PhysRev.101.453} {\bibfield  {journal}
  {\bibinfo  {journal} {Phys.Rev.}\ }\textbf {\bibinfo {volume} {101}},\
  \bibinfo {pages} {453} (\bibinfo {year} {1956})}\BibitemShut {NoStop}%
\bibitem [{\citenamefont {Jackura}\ \emph {et~al.}(2018)\citenamefont {Jackura}
  \emph {et~al.}}]{JPAC:2017dbi}%
  \BibitemOpen
  \bibfield  {author} {\bibinfo {author} {\bibfnamefont {A.}~\bibnamefont
  {Jackura}} \emph {et~al.} (\bibinfo {collaboration} {COMPASS and JPAC}),\
  }\href {\doibase 10.1016/j.physletb.2018.01.017} {\bibfield  {journal}
  {\bibinfo  {journal} {Phys.Lett.}\ }\textbf {\bibinfo {volume} {B779}},\
  \bibinfo {pages} {464–472} (\bibinfo {year} {2018})},\ \Eprint
  {http://arxiv.org/abs/1707.02848} {arXiv:1707.02848 [hep-ph]} \BibitemShut
  {NoStop}%
\end{thebibliography}

%

\end{document}